\documentclass{article}

\usepackage{PRIMEarxiv}

\usepackage[utf8]{inputenc} 
\usepackage{graphicx}
\usepackage{subcaption}
\usepackage{float}
\usepackage[colorlinks=true,linkcolor=blue,citecolor=blue]{hyperref}

\usepackage{times}
\usepackage{fullpage}
\usepackage{graphicx}
\usepackage{stmaryrd}
\usepackage{tikz}
\usepackage{doi}
\usepackage[fleqn]{amsmath}
\usepackage{amsthm}
\usepackage{amssymb}
\usepackage{mathrsfs}
\usepackage{mathtools}
\usepackage[subnum]{cases}
\usepackage{tocloft}
\usepackage{caption}
\usepackage{longtable}
\usepackage{enumitem}
\usepackage{courier}
\usepackage{algorithm}
\usepackage{algpseudocode}
\usepackage{pifont}
\usepackage{color}
\usepackage{verbatim}
\usepackage{hyperref}
\usepackage[page]{appendix}
\usepackage[scr=boondoxo]{mathalfa}
\usepackage[section]{placeins}
\usepackage{booktabs}
\usepackage{subcaption}

\setlist{labelindent=12pt}

\usepackage{physics}

\renewcommand{\mathbf}{\boldsymbol}

\usepackage{float}
\usepackage{capt-of}
\usepackage{tabularx}
\newcount\Comments  
\usepackage{color}
\definecolor{darkgreen}{rgb}{0,0.5,0}
\definecolor{purple}{rgb}{1,0,1}
\definecolor{uglycolor}{rgb}{0.5,0.5,0}
\newcommand{\kibitz}[2]{\ifnum\Comments=0\textcolor{#1}{#2}\fi}

\usepackage{tikz}

\title{Recent Surge in Public Interest in Transportation: Sentiment Analysis of Baidu Apollo Go Using Weibo Data
\thanks{\textit{\underline{Citation}}: 
\textbf{Wang, et al. Recent Surge in Public Interest in Transportation: Sentiment Analysis of Baidu Apollo Go Using Weibo Data.}} 
}

\author{
 {Shiqi Wang}\\
 School of GeoSciences\\
 University of Edinburgh\\
  \texttt{qtec@outlook.com}\\
   \And
  {Zhouye Zhao}\\
  College of Architecture and Urban Planning\\
  Tongji University\\
  \texttt{zhaozhouye@tongji.edu.cn}\\
     \And
  \textbf{Yuhang Xie}\\
  School of Social Science\\
  The University of Manchester\\
  \texttt{yuhang.xie-3@postgrad.manchester.ac.uk}\\
     \And
  \textbf{Mingchuan Ma}\\
  Tandon School of Engineering \\
  New York University\\
  \texttt{mm13145@nyu.edu}\\
       \And
  \textbf{Zirui Chen}\\
  Guangzhou Urban Planning \& Design Survey Research Institute\\
  \texttt{zenithchenr@gmail.com}\\
       \And
  \textbf{Zeyu Wang}\\
  Department of Urban Design and Planning\\ University of Washington \\
  \texttt{zeyuw1@uw.edu}\\
       \And
    \textbf{Bohao Su}\\
  Sichuan Highway Planning, Survey, Design and Research Institute Ltd\\
  \texttt{stephenbaihaouk@gmail.com}\\
  \And
    \textbf{Wenrui Xu}\\
  School of Architecture\\
  Tsinghua University\\
  \texttt{wenruixu@outlook.com}\\
  \And
    \And
  \textbf{Tianyi Li}\\
  Department of Civil Engineering\\
  Saint Louis University\\
  \texttt{tianyi.li.1@slu.edu}\\
}

\begin{document}
\maketitle

\begin{abstract}
Urban mobility and transportation systems have been profoundly transformed by the advancement of autonomous vehicle technologies. Baidu Apollo Go, a pioneer robotaxi service from the Chinese tech giant Baidu, has recently been widely deployed in major cities like Beijing and Wuhan, sparking increased conversation and offering a glimpse into the future of urban mobility.

This study investigates public attitudes towards Apollo Go across China using Sentiment Analysis with a hybrid BERT model on 36,096 Weibo posts from January to July 2024. The analysis shows that 89.56\% of posts related to Apollo Go are clustered in July. From January to July, public sentiment was mostly positive, but negative comments began to rise after it became a hot topic on July 21.

Spatial analysis indicates a strong correlation between provinces with high discussion intensity and those where Apollo Go operates. Initially, Hubei and Guangdong dominated online posting volume, but by July, Guangdong, Beijing, and international regions had overtaken Hubei. Attitudes varied significantly among provinces, with Xinjiang and Qinghai showing optimism and Tibet and Gansu expressing concerns about the impact on traditional taxi services.

Sentiment analysis revealed that positive comments focused on technology applications and personal experiences, while negative comments centered on job displacement and safety concerns. In summary, this study highlights the divergence in public perceptions of autonomous ride-hailing services, providing valuable insights for planners, policymakers, and service providers.
The model is published on Hugging Face at \url{https://huggingface.co/wsqstar/bert-finetuned-weibo-luobokuaipao} and the repository on GitHub at \url{https://github.com/GIStudio/trb2024}.

\end{abstract}

\section{Introduction}\label{section1}


The development of autonomous vehicle technology has significantly transformed urban mobility and transportation systems. At the forefront of these innovations is Apollo Go~\cite{apollo}, Baidu's autonomous ride-hailing service, which has seen substantial deployment in various Chinese cities. Apollo Go showcases the capabilities of autonomous driving technology and highlights its potential for improving urban mobility by reducing traffic congestion and enhancing transportation efficiency. With continuous advancements in sensing, automation, and computing technologies, autonomous driving is becoming increasingly accessible. It is expected to revolutionize future transportation systems by enhancing traffic safety~\cite{ye2019evaluating}, smoothing traffic flow~\cite{wang2022optimal}, and improving transportation energy efficiency~\cite{sun2022energy,li2023exploring}. While many of the potential benefits of autonomous vehicles (AVs) depend on achieving a high market penetration rate (MPR)~\cite{fagnant2015preparing}, recent studies indicate that even a modest MPR of intelligently controlled AVs can significantly enhance traffic flow~\cite{cui2017stabilizing}. Recently, Baidu Apollo Go has been widely deployed as a robotaxi service in major cities like Beijing and Wuhan, offering a glimpse into the future of urban mobility.


According to recent studies~\cite{schaller2018second,li2022taxi}, the deployment of autonomous ride-hailing services like Apollo Go has led to a paradigm shift in the urban transportation landscape. These services offer an alternative to traditional taxis and ride-hailing companies, aiming to provide safer, more reliable, and cost-effective transportation options. Wang et al.~\cite{wang2024competition} found that transferring autonomous vehicle technology between autonomous and traditional ride-hailing platforms can create a win-win scenario with higher profits for both platforms. While the integration of autonomous vehicles into the transportation network presents numerous benefits, it also introduces new challenges and considerations that need to be addressed. The public generally holds a cautious attitude towards autonomous taxis, as technological advancements and their reliability and functionality encourage trust, but concerns about potential job loss and the technology being dehumanizing foster negative sentiment~\cite{tussyadiah2017attitudes}. Even before the advent of autonomous ride-hailing services, new mobility services offered by TNCs, such as Uber and Lyft, had already dominated the market and changed transportation systems. However, while TNCs are successful in providing more choices for travelers, not all the impacts associated with TNCs are beneficial. For example, Komanduri et al.~\cite{komanduri2018assessing} show that the competitiveness of the TNC provider RideAustin versus transit has negative impacts on public transportation. Clewlow et al.~\cite{clewlow2017disruptive} studied the capacity utilization rate of taxis and TNCs in five major U.S. cities (Boston, Los Angeles, New York, San Francisco, and Seattle) in terms of time utilization rate and mileage utilization rate. The findings show that almost half of taxi and TNC operation time is not utilized. Inefficient operation lowers profit and leads to more congestion and emissions in urban cores where taxis are cruising for the next ride. 

Public opinion on using shared autonomous transportation in everyday life is quite divided. On the one hand, the safer people perceive autonomous vehicles to be, the more likely they are to use them, as they reduce the risk of collisions caused by traffic violations and human errors~\cite{montoro2019perceived}. Research indicates that high-income tech-savvy men living in urban areas and experiencing greater traffic accidents are more interested in these new technologies and are willing to pay more for them~\cite{bansal2016assessing}. Integrating autonomous driving with traditional public transportation can also improve mobility for the elderly and disabled~\cite{portouli2017public}. On the other hand, there is no consensus on the benefits of autonomous driving. Around half of the people express concerns about safety issues related to traffic accidents, robbery, and hacking~\cite{roche2019public}. Additionally, there are worries about the safety of autonomous vehicles operating alongside pedestrians and traditional vehicles in complex urban environments~\cite{battistini2020users}, as well as negative feelings about potential job losses and employment-related social exclusions related to automation~\cite{nikitas2021autonomous}. Currently, Apollo Go, China's first shared autonomous driving service platform operating in multiple cities, is scaling up its passenger testing operations. Gaining insights into public interest and perception is crucial for improving the platform's services.






With the rapid development of internet technology, social media platforms such as Twitter, Facebook, LinkedIn, Weibo, and others have become important channels for people to share their experiences and exchange views. As a carrier of public opinion, social media offers researchers a wealth of perspectives and thoughts due to its immediacy and diversity. Chinese social media platforms such as Weibo often include the location and time of the post when users publish content, associating these posts with specific places or events. Additionally, the tags included in the posts facilitate thematic categorization. Moreover, the content of the posts can reflect the cognition, emotions, and attitudes of the individuals posting them~\cite{10.1093/oxfordhb/9780199730018.013.0021,chen2023sentiment,tori2024performing}. These cognitive, emotional, and attitudinal factors significantly influence people's choices of transportation modes~\cite{parkany2004attitudes}. From a statistical perspective, analyzing the content of posts or tweets on social media to extract collective cognition, emotions, and attitudes can help identify potential factors influencing people's choices regarding autonomous transportation.

Recent developments in AI technology can be helpful in processing such massive text data for the analysis of transportation systems. Deep learning~\cite{lecun2015deep} has demonstrated promising capabilities in explaining complex nonlinear relationships and outperforming traditional approaches in various tasks such as image classification~\cite{krizhevsky2012imagenet} and natural language processing~\cite{graves2013speech}. It has also become a popular tool in the transportation community. For example, Cui et al.~\cite{cui2018deep} used a Long Short-Term Memory (LSTM) neural network to predict network-wide traffic speed, while Li et al.~\cite{li2022taxi} forecasted taxi demand and optimized city-wide taxi operations using an LSTM neural network. Deep learning is not only utilized in macroscopic traffic phenomena~\cite{cui2018deep, li2022taxi} but also widely applied to microscopic traffic phenomena, such as driving behavior models~\cite{li2023detecting,li2024car}.

One of the techniques used on social media for studying emotions and attitudes is Sentiment Analysis~\cite{tori2024performing,chen2023sentiment,singh2022sentiment}. Sentiment analysis is a key application of Natural Language Processing (NLP) that relies on various NLP techniques, including lexicon-based approaches, machine learning, and deep learning techniques, to classify text into sentiment categories such as positive, negative, and neutral~\cite{gunasekaran2023exploring}. NLP is an essential field in computer science and artificial intelligence, focused on enabling computers to comprehend and interact with human language. The evolution of NLP has been remarkable, starting with rule-based methods as documented in~\cite{10.1145/365153.365168}, then shifting to statistical approaches as seen in~\cite{32278}, and finally embracing deep learning techniques highlighted in~\cite{kalchbrenner-etal-2014-convolutional}. In the current landscape, NLP has reached a new milestone characterized by the advent of large, pretrained language models. The BERT model~\cite{devlin_bert_2019} stands out as a prominent example of this new era in NLP.
Social media sentiment analysis can more efficiently identify emotional tendencies with the help of pretrained language models~\cite{pan2024application}. Furthermore, fine-tuning techniques preserve the pretrained knowledge of these large models while significantly reducing the amount of labeled data and training time required~\cite{dimitrios_christofidellis_accelerating_2023}.


Text on social media often contains metaphors, harsh data, and the coexistence of many possible meanings for individual words or phrases, posing significant challenges to semantic understanding. For example, BERT's exceptional capability to discern sentiments within COVID-19-related tweets, a proficiency that can be applied to analyze transportation discourse during the pandemic with enhanced precision~\cite{saha2022vader}. The use of BERT models for travel pattern classifiers to determine whether each tweet is related to certain travel patterns and conduct sentiment analysis is to understand changes in people's attitudes towards pattern selection during the pandemic~\cite{chen2023sentiment}. A hybrid model that fuses BERT with BiLSTM and BiGRU for sentiment analysis of airline-related tweets demonstrates BERT's practical utility in interpreting customer sentiment—a key asset for the transportation industry~\cite{talaat2023sentiment}. In these studies, the preprocessing phase primarily entails text cleansing and normalization, tokenization, and length adjustment to ensure uniformity and proper formatting, with tailored refinements made to accommodate specific elements like URLs, mentions, topic tags, special characters, and extraneous content~\cite{palani2021t,saha2022vader,chen2023sentiment,talaat2023sentiment}. At present, the technology specifically used for sentiment analysis relies on established NLP preprocessing techniques and does not use effective data initialization and preprocessing techniques, which will seriously affect the performance of sentiment analysis models~\cite{bordoloi2023sentiment}.

For robotaxis, there are certainly benefits in energy utilization~\cite{zhou2023robotaxi}, such as low carbon development, but at the same time, people are concerned about their safety, and crashes have already happened~\cite{koopman2024lessons}. Since Apollo Go has just been deployed in China and gained popularity early this year, and we do not have access to its trajectory data and OD data, it is not possible to study explicitly how it will influence the current traffic patterns. However, we are interested in understanding how the general public perceived this revolutionary service before it was widely deployed citywide and in other cities. We use crawled Weibo data obtained via API to study public opinion on this revolutionary service in China.


The goal of this study is to provide a scenario to observe how the general public reacts to the rapid and large-scale deployment of robotaxis using real-world data from Weibo. Specifically, this paper aims to answer the following three questions:

\begin{enumerate}[leftmargin = 12pt,labelindent=12pt]

\item How can we use natural language data to better understand new mobility services?

\item What is the general public's emotional reaction to the sudden appearance of the robotaxi?

\item What suggestions and insights can be informed to the public and transportation agencies?

\end{enumerate}

The remainder of this article is outlined as follows. First, we provide a review of the relevant literature on mobility services and the methodologies used in this study. Second, we give a detailed summary of the experimental data, including data collection, processing, and the study sites. Third, we implement the proposed methods and present the results of our analysis. Finally, we conclude with a discussion of the findings from this study and directions for future research.

\section{Experimental Data}
The dataset consists of collected opinions and comments on Weibo regarding the Baidu Apollo Go service. Due to the difficulty in identifying the information on the posters and concerns regarding online privacy protection, there are issues such as ambiguous spatial locations and varying quality of comments. This presents challenges for data cleaning and analysis. However, these data include all topic data from January to July 2024, which is sufficient to conduct sentiment analysis on public opinion using natural language processing. The 'Content' column contains over 30,000 public views posted by Weibo users located in different regions across China. Therefore, the dataset spans a total of six months before and after the peak period, with data publication locations covering all provinces, municipalities, and autonomous regions in China, which is sufficient for sentiment analysis.

\begin{figure}
    \centering
    \includegraphics[width=0.6\linewidth]{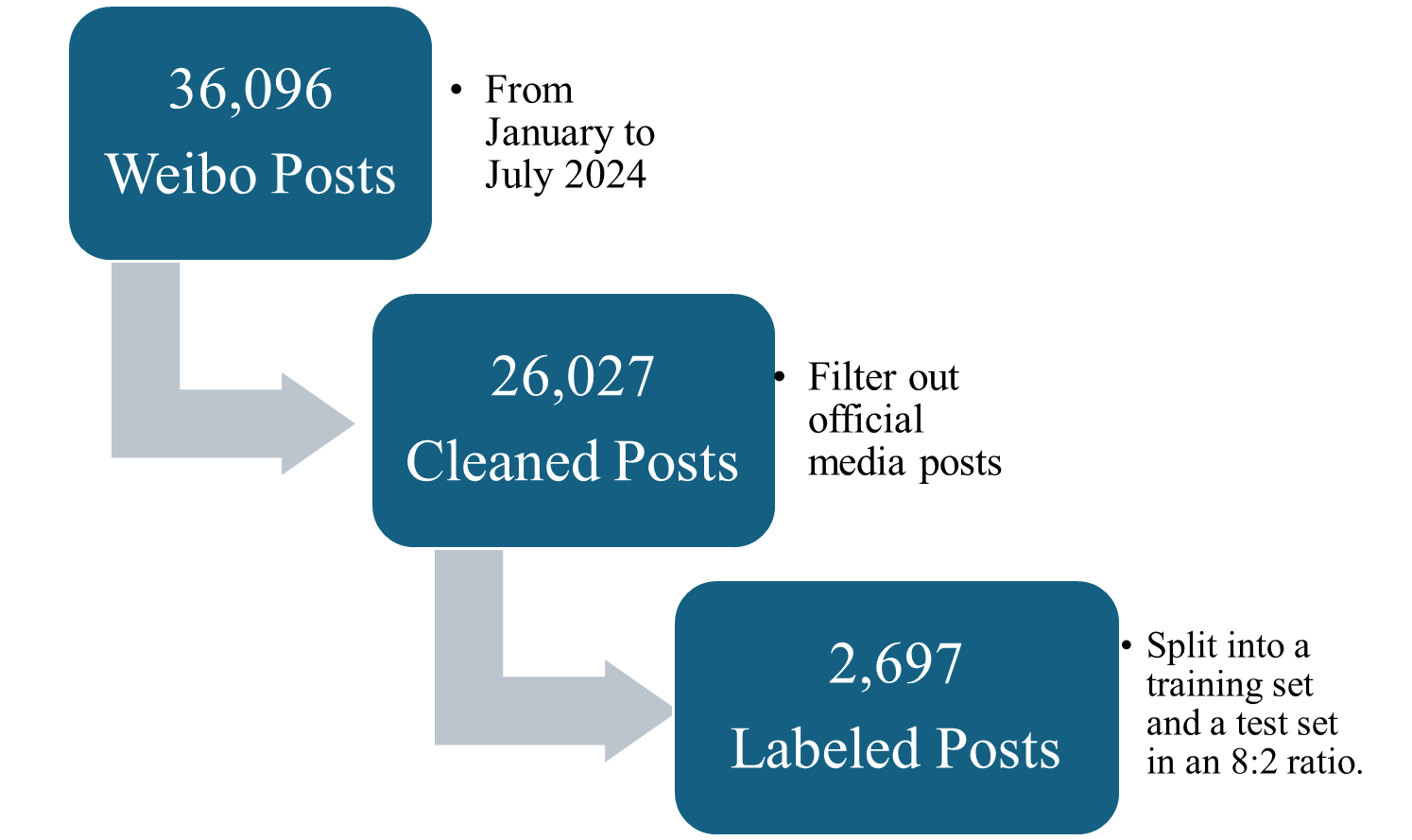}
    \caption{Description of the amount of data used in each phase.}
    \label{fig:data-process}
\end{figure}

By using the Weibo API service, 36,096 Weibo posts and comments containing the ``Baidu Apollo Go" tag from January to July 2024 were collected (Figure~\ref{fig:data-process}). The data contains user name, IP location, time, attitude count, content, and other relevant information. A sample dataset is shown in Table~\ref{tab:sample}. The username column uniquely identifies each user, the content column records the text or emoji content posted by each user, and the location column records the IP address from which each comment was posted. For privacy protection, the IP locations for each record are only accurate at the province or municipality level. As discussed above, these data will be used to analyze people's attitudes toward the Baidu Apollo Go service. The content column contains user comments or experiences with the Baidu Apollo Go service.

\begin{table}[h]
    \centering
    \caption{Data Structure of Selected Variables.}
     \scalebox{0.71}{
    \begin{tabular}{l p{4cm}c c c c c}
     \toprule
    \bf User ID &\bf Content & \bf Reposts count & \bf Attitudes count &\bf Comments count & IP location & \bf Time \\
    \hline
    2800847042 & Direct to Tianhe Airport! Baidu Apollo Go autonomous vehicles take you home for the Spring Festival travel season. & 5 & 33 & 12 & Post in Hubei & 2024/1/26 17:54 \\
    \hline
    6372873842 & Autonomous taxis cross the Yangtze River BaiduApolloGo begins cross-river operations. & 20 & 56 & 5 & Post in Hubei & 2024/2/27 19:44 \\
    \hline
    5493776217 & Immersive experience with autonomous vehicles, super cool! & 12 & 22 & 13 & Post in Beijing & 2024/3/23 19:02 \\
    \hline
    7821592882 & Cars can drive themselves now! It feels very convenient. & 1 & 5 & 1 & Post in Guangdong & 2024/4/15 18:06 \\
    \hline
    5953250708 & Fully support Baidu Apollo Go! No more dealing with taxi drivers rolling their eyes and complaining about short trips. & 16 & 28 & 3 & Post in Anhui & 2024/7/11 18:55 \\
    \hline
    3735148384 & Be more tolerant of autonomous vehicles. Support Baidu Apollo Go and China's autonomous driving industry. & 1 & 1 & 0 & Post in Fujian & 2024/7/13 18:57 \\
			\bottomrule
    \end{tabular}}
    \label{tab:sample}
\end{table}


\subsection{Data Cleaning}
The original dataset contains two types of irrelevant content: first, information tagged with ``Baidu Apollo Go" but unrelated in content; second, posts by official media and self-media accounts that lack a clear attitude. Before processing the data, these irrelevant posts need to be removed. A list of official media was used to match the username of each post, and a set of keywords was used for regex search. Records posted by these accounts were dropped directly due to their lack of clear attitude as their content focused on describing facts.


After removing official information, unrelated content needed to be identified through manual annotation. Some users posted long texts, such as blogs recording their daily activities, where Baidu Apollo Go is mentioned but not the main focus. However, some users expressed their attitudes within a sentence in the long text. These records should be annotated to capture the attitude and not be dropped. In this study, 20 percent of long texts were manually annotated. Once the long texts are manually annotated, feature extraction methods can be applied to the remaining records


\subsection{Data Processing}


A total of nine individuals participated in the data annotation process using the open-source software doccano~\cite{doccano}, resulting in the annotation of 2,797 Weibo posts. After manually removing excessively long and meaningless texts, 2,697 entries were utilized for fine-tuning the large model. The dataset was divided into a training set and a validation set in a ratio of 4:1. These entries were categorized into four labels: ``Positive" with 1,089 samples, ``Neutral" with 659 samples, ``Negative" with 629 samples, and "Drop" with 320 samples. A summary of statistics of the different categories is presented in Table~\ref{tab:stat}.

\begin{table}[h]
\centering
\caption{Distribution of Labeled Weibo Posts.}
\begin{tabular}{cc}
\toprule
Category & Number of Samples \\
\hline
Positive & 1,089 \\
Neutral & 659 \\
Negative & 629 \\
Drop & 320 \\
\hline
Total & 2,697 \\
\bottomrule
\end{tabular}
\label{tab:stat}
\end{table}

After applying the sentiment analysis and classification algorithm, the attitudes were categorized into three classes: positive, negative, and neutral. The classification standard is based on the emotional attitude presented by the text data in the content column. Texts exhibiting emotions such as excitement, enthusiasm, support, and encouragement were classified as positive, indicating support and anticipation for Baidu Apollo Go and similar technological services, with the belief that further deep testing and early commercial operation are warranted. Texts expressing dissatisfaction, disgust, disapproval, or abusive language were classified as negative, indicating a negative attitude towards Baidu Apollo Go and similar services, suggesting that the development of this technology should not continue at this time. Other content, such as news articles, media reports, and financial information, were classified as neutral, as they merely stated objective facts without emotional bias. By combining the time and IP location records, we could determine the attitudes of people from different times and places.

\subsection{Study Site}
\begin{figure}[h]
    \centering
    \includegraphics[width=0.7\linewidth]{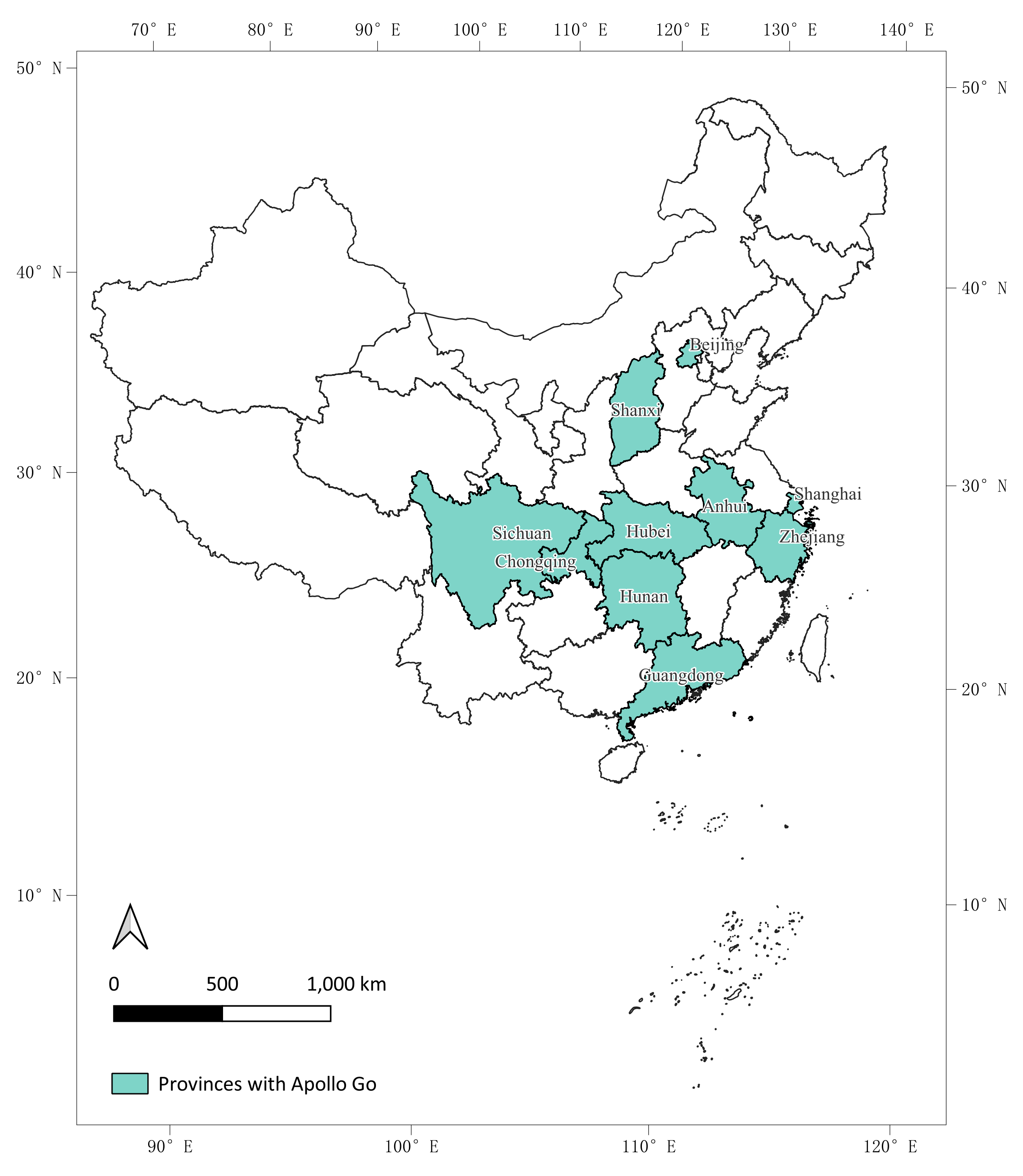}
    \caption{Operating conditions of Apollo Go in different regions.}
    \label{fig:provinces_with_apollo_go}
\end{figure}

Our research used social media data to analyze public attitudes towards Apollo Go across China's provinces, providing a comprehensive view of nationwide public sentiment. This includes provinces that already operate Apollo Go and those that do not. Apollo Go is currently operating in 11 cities within 10 provinces, as shown in Figure~\ref{fig:provinces_with_apollo_go} and Table~\ref{tab:cities}. The time column of Table~\ref{tab:cities} also shows the first time Apollo Go was launching regular commercial operations to the public. These provinces are known for their rapid economic growth and high levels of technological adoption, having achieved regular commercial operation of Apollo Go for two to four years. Studying these areas provides insights into public attitudes where autonomous vehicles are becoming more integrated into daily life. While focusing on the provinces where Apollo Go operates, our study also considers data from non-operational provinces. This broader perspective ensures that we capture public sentiment from regions with different levels of exposure to autonomous vehicle technology. This approach allows us to examine regional differences in acceptance, concerns, and overall sentiment towards Baidu's Apollo Go.


\begin{table}[h]
    \centering
    \caption{Operating Conditions of Apollo Go.~\cite{ApolloGO}}
    \begin{tabular}{ccc} 
    \toprule
         Province&    City& Time\\ \hline 
         Beijing& \
      Beijing&2/5/2021\\ 
 Hubei& Wuhan& 10/5/2022\\ 
 Hunan& Changsha& 19/4/2020\\ 
 Guangdong& Guangzhou& 17/7/2021\\ 
 Shanghai& Shanghai& 12/9/2021\\ 
 Guangdong& Shenzhen& 17/2/2022\\ 
 Shanxi& Yangquan& 27/2/2022\\ 
 Chongqing& Chongqing& 7/8/2022\\ 
 Sichuan& Chengdu& 24/7/2022\\ 
 Anhui& Hefei& 9/19/2022\\ 
 Zhejiang& Wuzhen& 03/26/2022\\ 
 \bottomrule
 \end{tabular}
    \label{tab:cities}

\end{table}

\section{Methodology}
\begin{figure}
    \centering
    \includegraphics[width=0.7\linewidth]{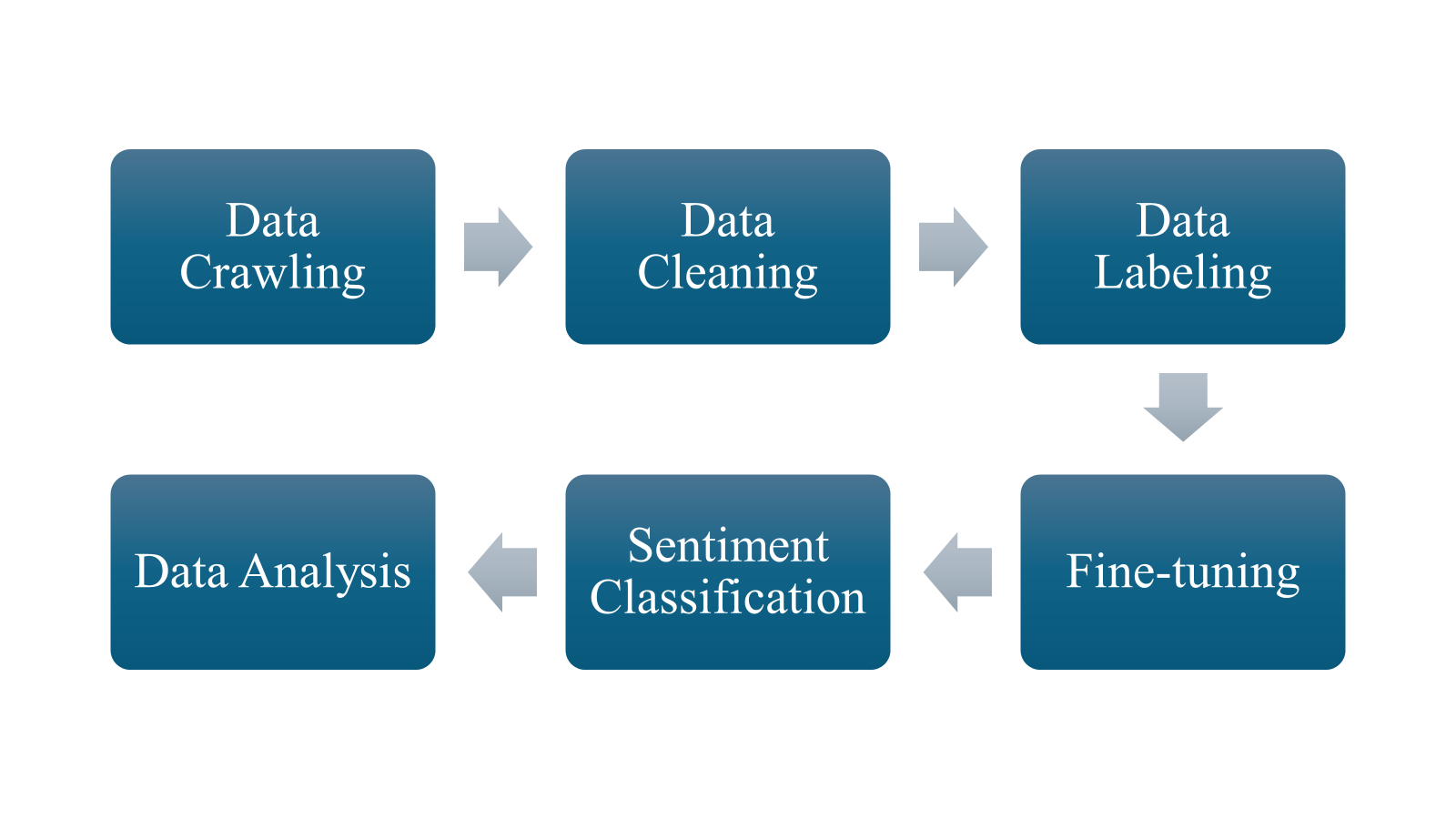}
    \caption{Outline of the methodology used in this study, including steps for data crawling, data cleaning, data labeling, fine-tuning, sentiment classification, and data analysis.}
    \label{fig:methodology}
\end{figure}

Our analysis pipeline can be described in Figure~\ref{fig:methodology}. In this study, we began by collecting a large set of Weibo posts related to Apollo Go. From this extensive dataset, we selected a sample of 2,697 posts for detailed labeling and fine-tuning. The first step involved meticulously labeling these selected posts. Posts that were deemed irrelevant to Apollo Go were labeled as 'Drop'. The remaining posts were categorized based on their sentiment into three groups: positive, neutral, and negative. This categorization allowed us to analyze the sentiment distribution and understand public perception more effectively.

To analyze the sentiment of all collected Weibo posts, we fine-tuned the 'bert-base-chinese' model~\cite{devlin_bert_2019}, which is pre-trained specifically for the Chinese language and available on Hugging Face. This model was chosen for its proven effectiveness in handling Chinese text, its ability to capture contextual nuances, and its strong performance in various natural language processing tasks. Fine-tuning the BERT model involved training it specifically on our labeled dataset of Weibo posts. 

The following hyperparameters were utilized during the training process: the learning rate was set to 1e-05, ensuring minimal updates during fine-tuning to make subtle adjustments to the pre-trained BERT model's parameters. This approach enables the model to adapt to our specific tasks without significantly disrupting the representations it has already learned. The training batch size and evaluation batch size were both set to 8, balancing computational efficiency and model performance. A seed value of 42 was used to ensure reproducibility. The Adam optimizer, configured with betas of (0.9, 0.999) and an epsilon value of 1e-08, was employed. The learning rate scheduler was of the linear type, and the model was trained over 5 epochs. This adaptation aimed to enhance the model's performance in sentiment classification of Apollo Go-related content.

After the fine-tuning process, the model achieved an accuracy of 0.59. This accuracy indicates the model's capability to correctly classify the sentiment of Weibo posts, demonstrating the feasibility of using BERT for sentiment analysis in this context. For further details on the model and fine-tuning process, you can refer to the model repository on Hugging Face at the following link: \url{https://huggingface.co/wsqstar/bert-finetuned-weibo-luobokuaipao} and the repository on GitHub: \url{https://github.com/GIStudio/trb2024}. After fine-tuning the model, we directly classified the initially cleaned Weibo data.


\section{Model Results \& Analysis}

After building our model, we describe the trends of Apollo Go both temporally and spatially. We focus on the period from January to June, during which Apollo Go had not yet gained significant popularity on the internet. Additionally, we analyzed another group from July, when Apollo Go became a trending topic online.

\subsection{Temporal Features Analysis}
\begin{figure}
    \centering
    \includegraphics[width=1\linewidth]{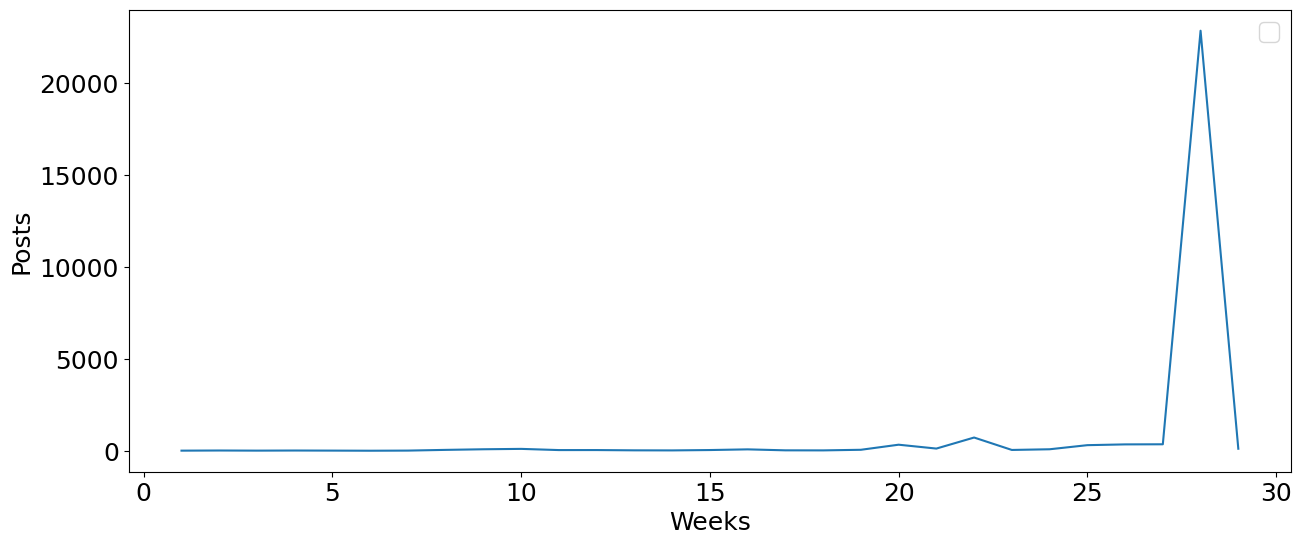}
    \caption{Weekly discussion trend from the year of 2024, with a focus on July which begins in the 27th week. Data for this analysis is collected up to July 14th.}
    \label{fig:weekly-discusstion-trend-from-the-year-of-2024}
\end{figure}

\begin{figure}
    \centering
    \begin{subfigure}[b]{1\linewidth}
        \centering
        \includegraphics[width=1\linewidth]{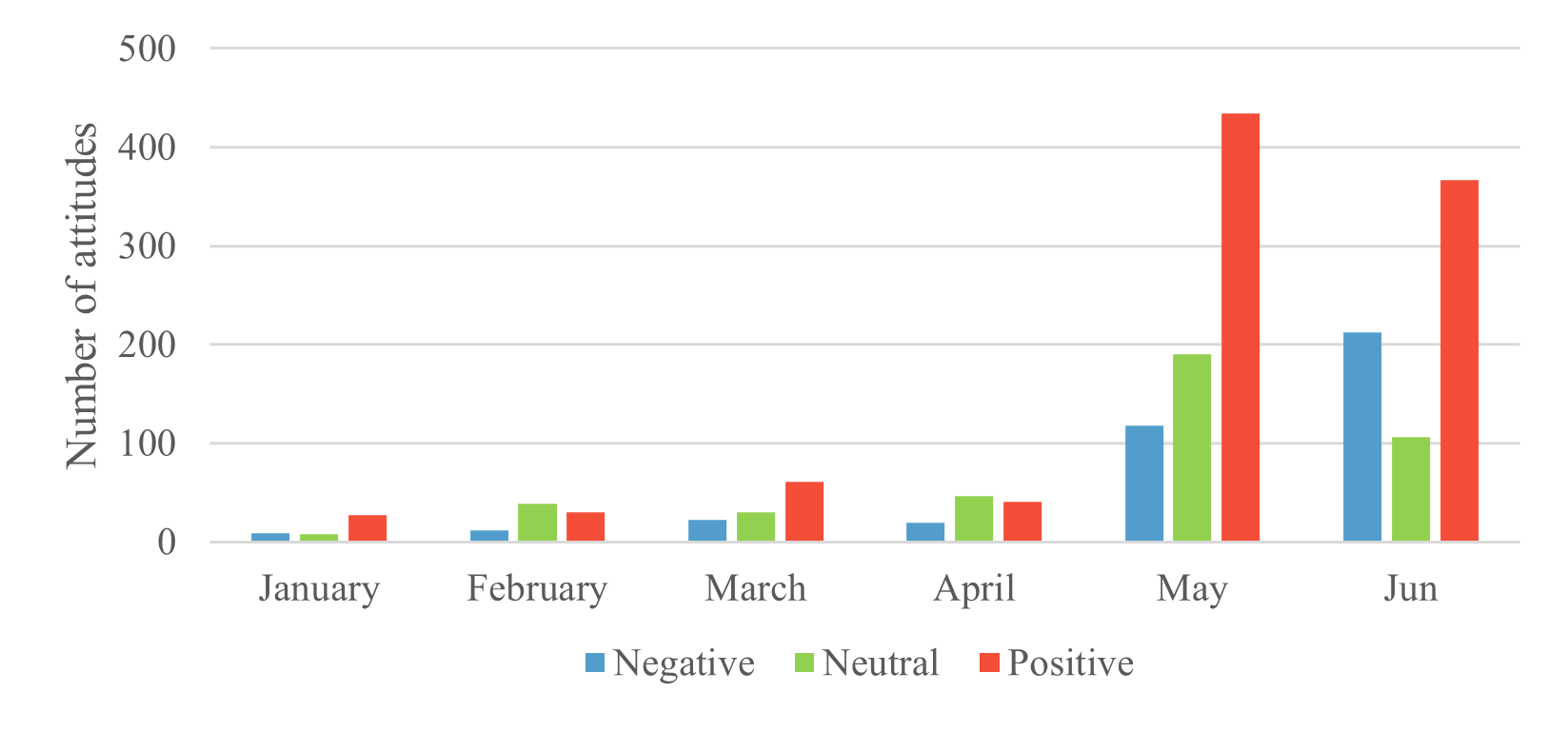}
        \caption{Pane1 1: Distribution of different attitudes per month from January to June 2024.}
        \label{fig:panel1-Number-of-different-attitudes-per-month-from-January-to-June}
    \end{subfigure}
    \vspace{1em} 
    \begin{subfigure}[b]{1\linewidth}
        \includegraphics[width=1\linewidth]{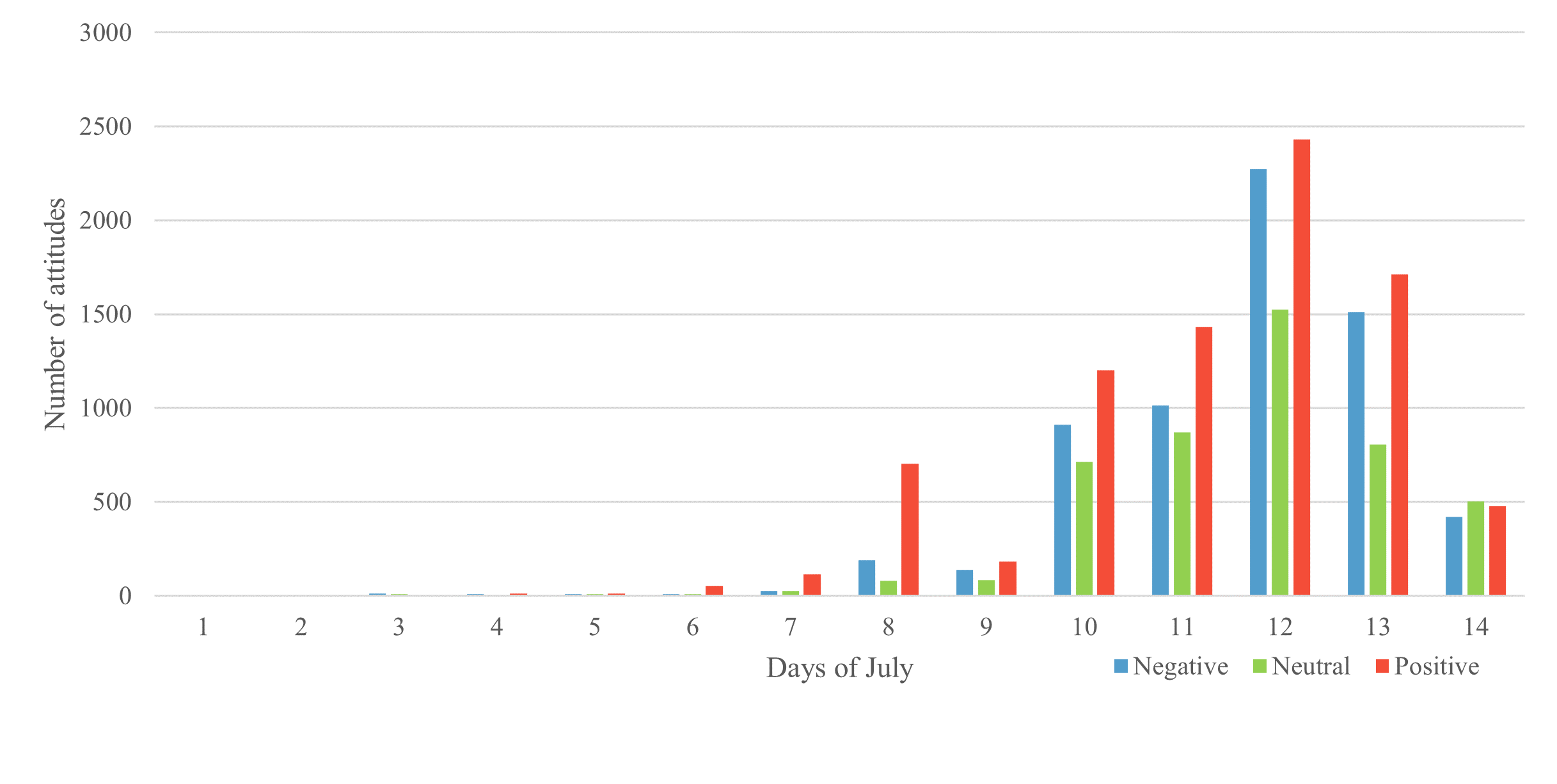}
        \caption{Pane1 1: Distribution of different attitudes per day in July 2024.}
        \label{fig:panel2-Number-of-different-attitudes-per-day-in-July}
    \end{subfigure}
    \caption{Distribution of different attitudes.}
    \label{fig:combined}
\end{figure}

After grouping the data by posts per month, we found that 89.56\% of posts related to Apollo Go are concentrated within July. As shown in Figure~\ref{fig:weekly-discusstion-trend-from-the-year-of-2024} and Table~\ref{tab:number-of-different-attitudes-per-month-from-1-to-7-2024}, this trend is clearly reflected in the data analysis. This matches news articles that Apollo Go was commercially operated in Wuhan, Hubei in July, and this topic became popular over the internet. Under this situation, we separated the analysis into two-time phases, January to June and July, to analyze the change in attitudes and the Spatial distribution of attitudes on Weibo. This analysis is detailed in Figure~\ref{fig:panel1-Number-of-different-attitudes-per-month-from-January-to-June}, Figure~\ref{fig:panel2-Number-of-different-attitudes-per-day-in-July} and Table~\ref{tab:number-of-different-attitudes-per-day-in-july-2024}.

From January to July, public attitudes towards Apollo Go were predominantly positive. In early July, "Apollo Go" gained significant public attention after trending on social media. On May 15, 2024, Baidu Apollo hosted the Apollo Day 2024 event in Wuhan, which showcased the latest autonomous driving foundation model, generating significant media coverage and public interest. The discussion around the service surged from July 8th, reaching a peak on July 12th, before gradually declining. Enthusiasm for this service continues to grow due to the novelty of the technology and perceived improvements in ride-hailing quality and convenience. However, from July onwards, as the volume of public discourse increased, the proportion of negative comments began to rise. Concerns about job displacement due to the automation of driving roles became more pronounced. Additionally, some users expressed dissatisfaction with the presence of safety operators in the vehicles, feeling that it undermined the autonomy of the service. This shift in sentiment highlights the complex characteristics of public attitudes towards autonomous vehicles. While there is considerable enthusiasm for technological advancements and potential benefits, there are also significant concerns and resistance, particularly related to socioeconomic impacts and the perceived authenticity of the autonomous experience. 


\begin{table}
    \centering
      \caption{Number of different attitudes per month from January to July 2024.}
    \begin{tabular}{lllll}\hline
      
        Month& Negative & Neutral & Positive & Total \\ \hline
        January & 9  & 8  & 27  & 44 \\ 
        February & 12 & 39 & 30 & 81 \\ 
        March & 22 & 30 & 61 & 113 \\ 
        April & 19 & 46 & 41 & 106 \\ 
        May & 118 & 190 & 434 & 742 \\ 
        June& 212 & 106 & 367 & 685 \\
         July& 6552& 4670& 8375&19597\\ \hline
    \end{tabular}
    \label{tab:number-of-different-attitudes-per-month-from-1-to-7-2024}
\end{table}

\begin{table}
    \centering
    \caption{Number of different attitudes per day in July 2024.}
    \begin{tabular}{lllll}\hline
    
        Date& Negative & Neutral & Positive & Total \\ \hline
        1/7& 2 & 1 & 4 & 7 \\ 
        2/7 & 3 & 1 & 3 & 7 \\ 
        3/7 & 12 & 8 & 5 & 25 \\ 
        4/7 & 9 & 5 & 11 & 25 \\ 
        5/7 & 10 & 9 & 13 & 32 \\ 
        6/7 & 8 & 7 & 54 & 69 \\ 
        7/7 & 25 & 25 & 115 & 165 \\ 
        8/7 & 188 & 80 & 702 & 970 \\ 
        9/7 & 139 & 82 & 183 & 404 \\ 
        10/7 & 910 & 715 & 1201 & 2826 \\ 
        11/7 & 1013 & 870 & 1434 & 3317 \\ 
        12/7 & 2273 & 1524 & 2429 & 6226 \\ 
        13/7 & 1511 & 805 & 1711 & 4027 \\ 
        14/7 & 422 & 503 & 480 & 1405 \\ \hline
    \end{tabular}

    \label{tab:number-of-different-attitudes-per-day-in-july-2024}
\end{table}

\begin{figure}
    \centering
    \includegraphics[width=1\linewidth]{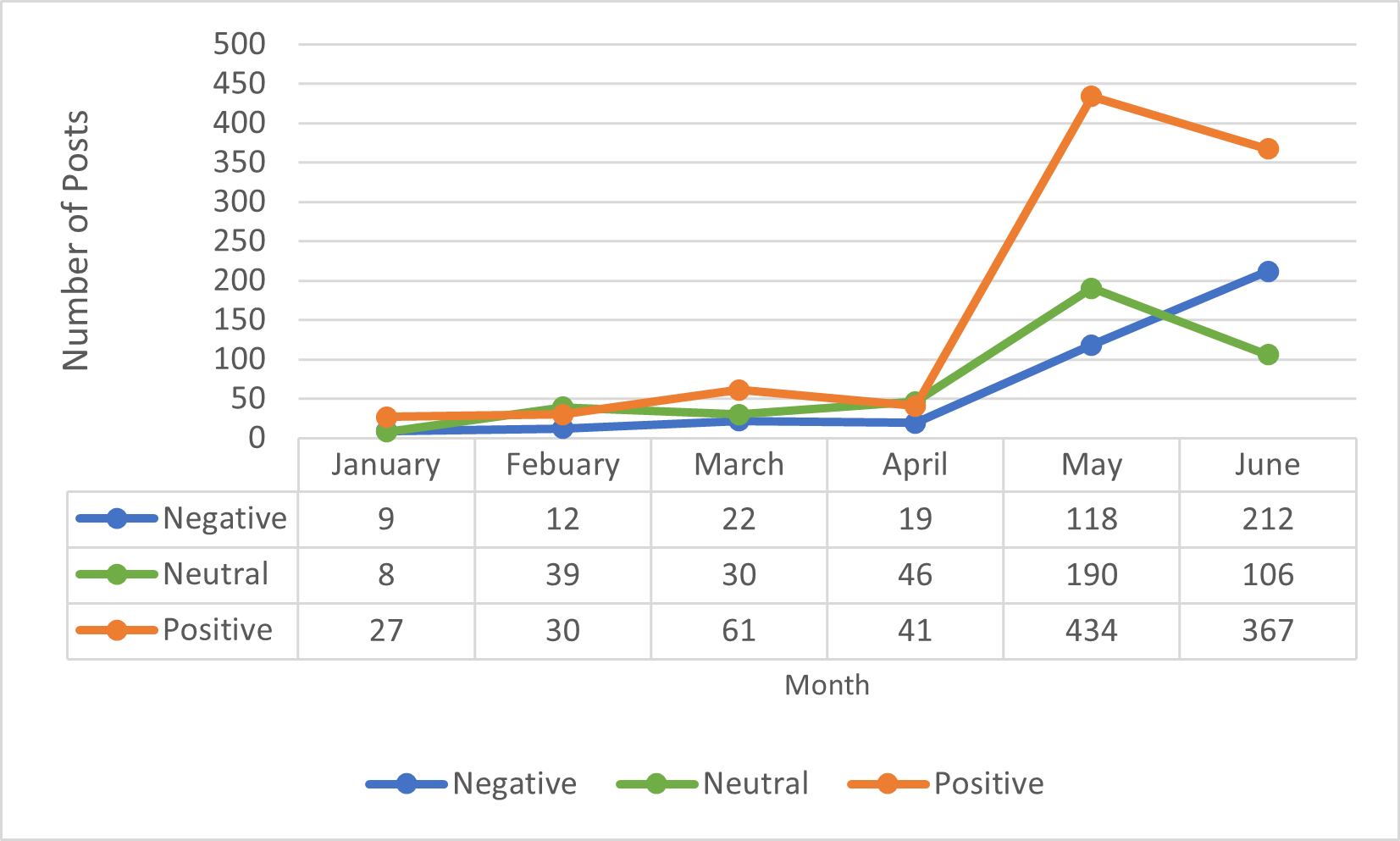}
    \caption{The trend of attitudes from January to June.}
    \label{fig:enter-label_1}
\end{figure}
\begin{figure}
    \centering
    \includegraphics[width=1\linewidth]{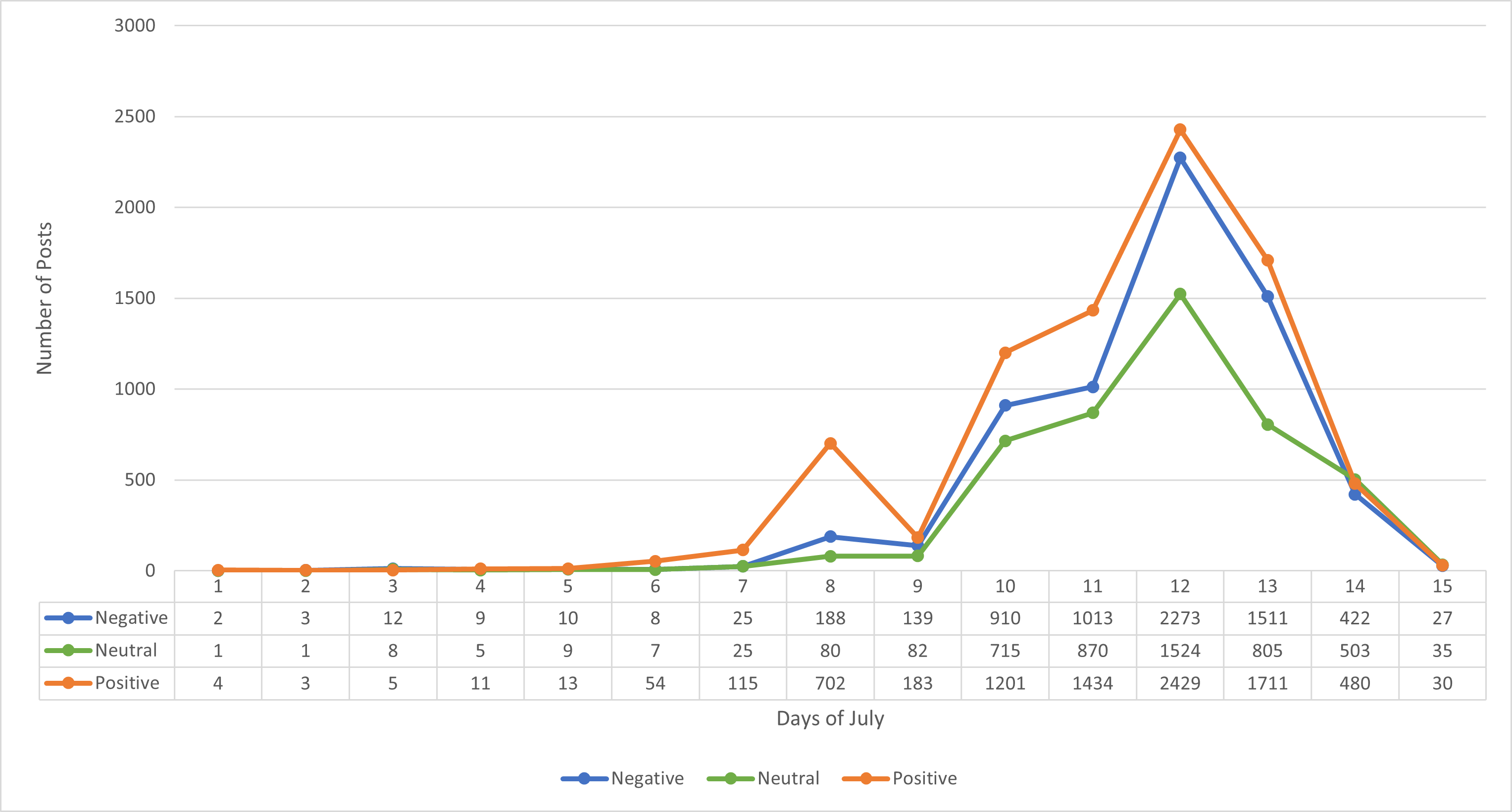}
    \caption{The trend of attitudes in July.}
    \label{fig:enter-label_2}
\end{figure}

\subsection{Spatial Features Analysis}
\begin{figure}
    \centering
    \begin{subfigure}[b]{1\linewidth}
        \centering
        \includegraphics[width=1\linewidth]{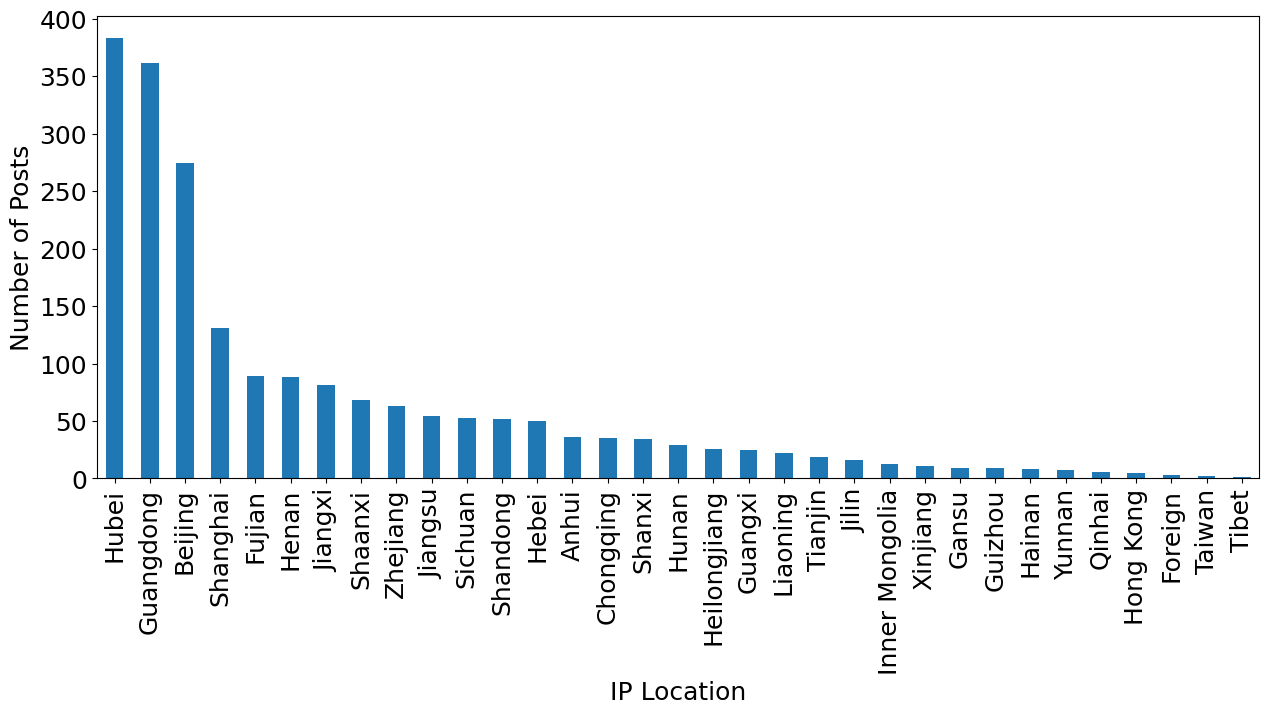}
        \caption{Panel 1: Posts from January to June 2024.}
        \label{fig:posts_jan_to_jun}
    \end{subfigure}
    \vspace{1em} 
    \begin{subfigure}[b]{1\linewidth}
        \centering
        \includegraphics[width=1\linewidth]{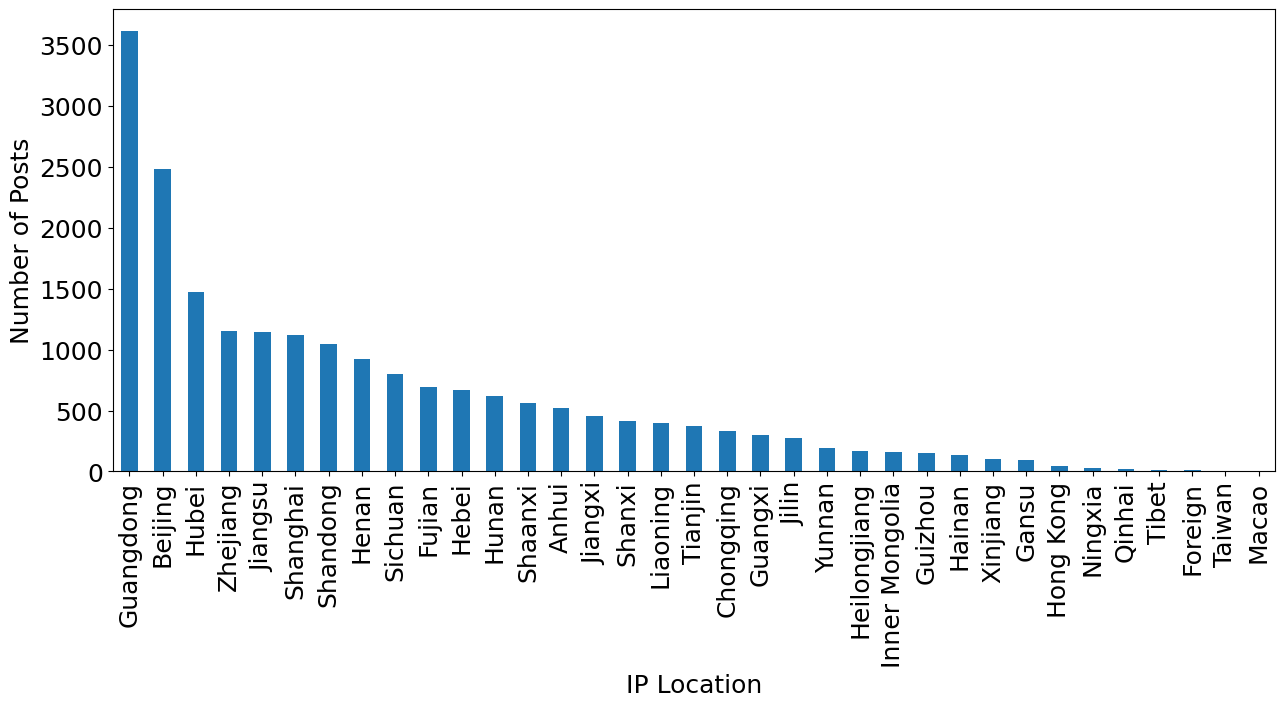}
        \caption{Panel 2: Posts in July 2024.}
        \label{fig:posts_july}
    \end{subfigure}
    \caption{Posts over different time periods.}
    \label{fig:combined_posts}
\end{figure}

In the first 6 months, Hubei and Guangdong provinces dominated online posting volume in China, as shown in Table~\ref{tab:Posts in 2024} and Figure~\ref{fig:combined_posts}. Beijing and Shanghai have also ranked high on the discussion board tables. Most of the relevant conversations came from these regions making them all top four dispensers for discourse during the first few months of 2024. Comparing July with the initial six months, it seems that Apollo Go has enjoyed robust adoption in the Chinese mega-cities. Guangdong and Beijing significantly increased postings in July, and most of the top 10 cities~\ref{tab:Posts in 2024} increased posts nearly tenfold in July. However, Hubei didn't increase as much compared to other provinces. The level of discussion abroad has also surpassed that in Wuhan. This might be caused by the discussion in Hubei becoming localized and primarily happening offline.



\begin{table}
    \centering
    \caption{Post amounts in 2024.}
    \begin{tabular}{ccccc}
    \toprule
 & \multicolumn{2}{c}{January to June} & \multicolumn{2}{c}{July}\\
          &IP Location& Post Counts & IP Location&Post Counts \\
          \hline
          1&Hubei& 383& Guangdong
&3615\\
          2&Guangdong& 362& Beijing
&2483\\
          3&Beijing& 275& Hubei
&1477\\
          4&Shanghai& 131& Zhejiang
&1155\\
          5&Fujian& 89& Jiangsu
&1147\\
          6&Henan& 88& Shanghai
&1120\\
          7&Jiangxi& 81& Shandong
&1051\\
          8&Shanxi& 68& Henan
&927\\
  9&Zhejiang&63& Sichuan&798\\
 10& Jiangsu& 54& Fujian&698\\
 \bottomrule
    \end{tabular}

    \label{tab:Posts in 2024}
\end{table}

\begin{figure}
    \centering
    \includegraphics[width=1\linewidth]{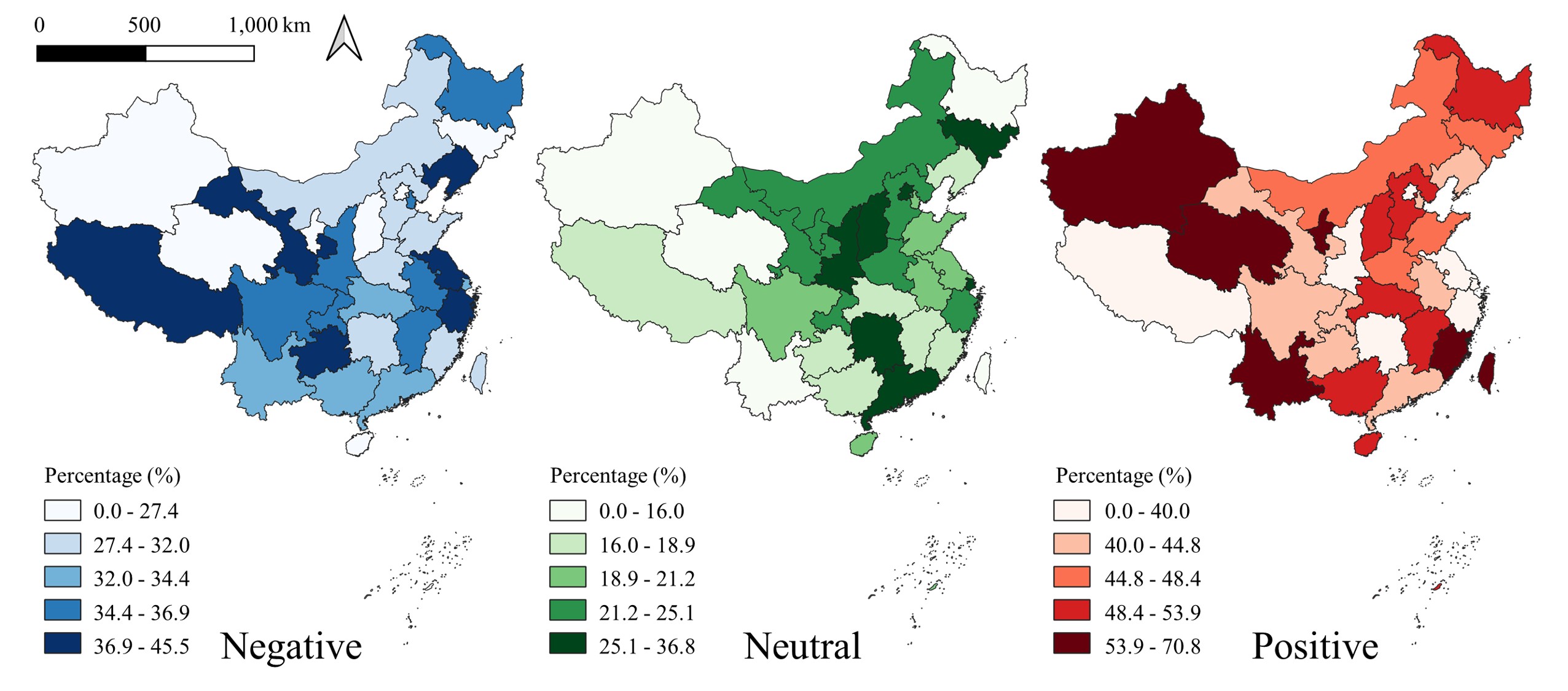}
    \caption{Spatial distribution of the number of negative, neutral, and positive attitudes from January to July 2024.}
    \label{fig:Jan to Jul number}
\end{figure}

\begin{figure}
    \centering
    \includegraphics[width=1\linewidth]{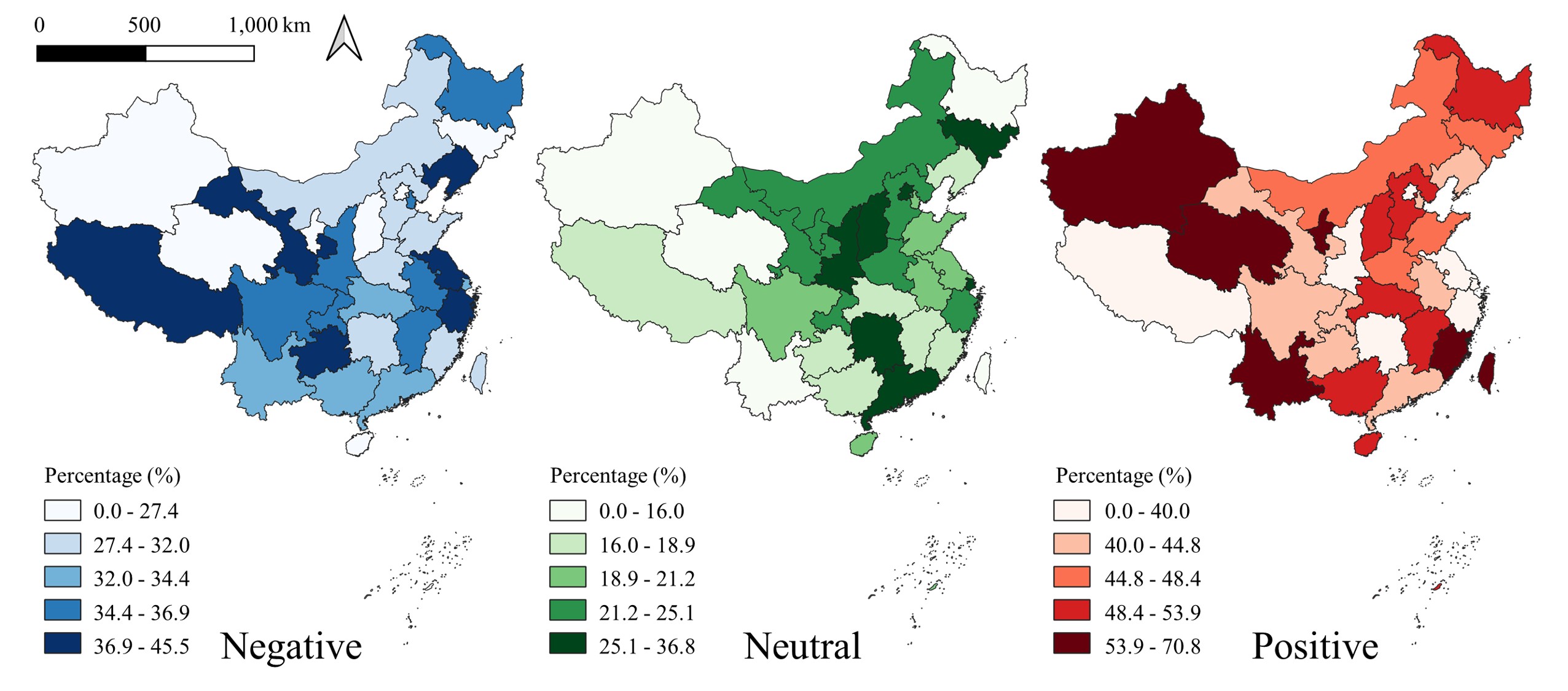}
    \caption{Spatial distribution of the percentage of negative, neutral, and positive attitudes from January to July 2024.}
    \label{fig:Jan to Jul percentage}
\end{figure}

The spatial distribution of public attitudes towards the Apollo Go from January to July was analyzed. Figure~\ref{fig:Jan to Jul number} depicts the distribution of the number of each attitude category across various regions, while Figure~\ref{fig:Jan to Jul percentage} illustrates the corresponding percentage distribution. Both maps utilize a color gradient to convey the proportion of attitudes: negative attitudes are represented in blue, neutral attitudes in green, and positive attitudes in red.

The eastern coastal provinces show a higher level of discussion intensity regarding Apollo Go, particularly in Beijing, Guangdong, and Shandong. Public concerns predominantly focus on employment issues and the perceived irreplaceability of human work in the era of artificial intelligence. There is a strong correlation between the provinces with high discussion intensity and those where Apollo Go operates. Some individuals express excitement due to Apollo Go's competitive pricing compared to traditional ride-hailing services. Conversely, interest in Apollo Go is relatively lower in western regions, and attitudes vary significantly among provinces. Xinjiang and Qinghai provinces are predominantly shaded in deep red, indicating optimism about the unstoppable trend of autonomous driving and anticipation for its development. Tibet and Gansu provinces are primarily shaded in deep blue, reflecting concerns about the impact of autonomous vehicles on traditional taxi and ride-hailing services.

From 8 July to 14 July, the spatial characteristics of panel data reveal several trends. In terms of quantity, as shown in Figure~\ref{fig:8 July to 14 July number}, there is a general trend of high attention across coastal provinces, even in regions where Apollo Go does not operate. This suggests a pervasive interest in autonomous ride-hailing services, possibly influenced by media coverage and broader technological discourse. Specifically, Xinjiang stands out with exceptionally high attention compared to other western and northern provinces, indicating a unique regional interest in the topic.

In terms of percentages, as depicted in Figure~\ref{fig:8 July to 14 July percentage}, the analysis focuses on the period from July 10 to 14 to discuss public attitude changes, as initial days show lower data volume, making them less representative of public attitudes. Overall, there is an expansion in the coverage of positive attitudes and a reduction in the coverage of negative attitudes. The percentage of neutral attitudes is evenly distributed across provinces, suggesting widespread public awareness of Apollo Go facilitated by news and other social media content over time. Provinces with high values of positive attitudes typically align with provinces where there are also high values of neutral attitudes. This alignment often corresponds to regions where there has been extensive media coverage of Apollo Go's operations or advancements in autonomous driving technology. This indicates that media coverage plays a crucial role in shaping public attitudes towards Apollo Go. News reports effectively highlight the benefits of autonomous ride-hailing services such as convenience and technological advancements, thereby mitigating public concerns surrounding the technology.

\begin{figure}
    \centering
    \begin{subfigure}[b]{1\linewidth}
        \centering
        \includegraphics[width=0.9\linewidth]{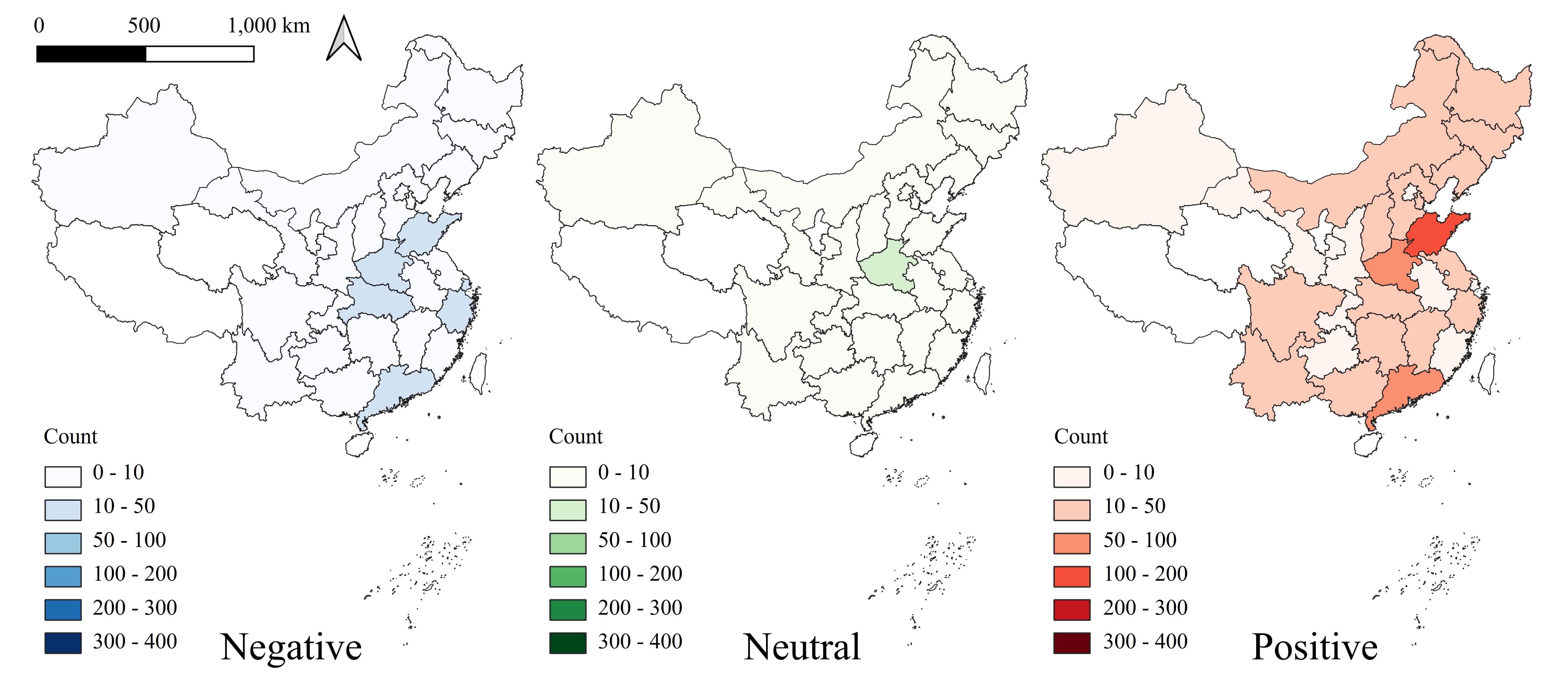}
        \caption{July 8, 2024}
        \label{fig:count_07-08}
    \end{subfigure}
    \vspace{1em} 
    \begin{subfigure}[b]{0.9\linewidth}
        \centering
        \includegraphics[width=0.9\linewidth]{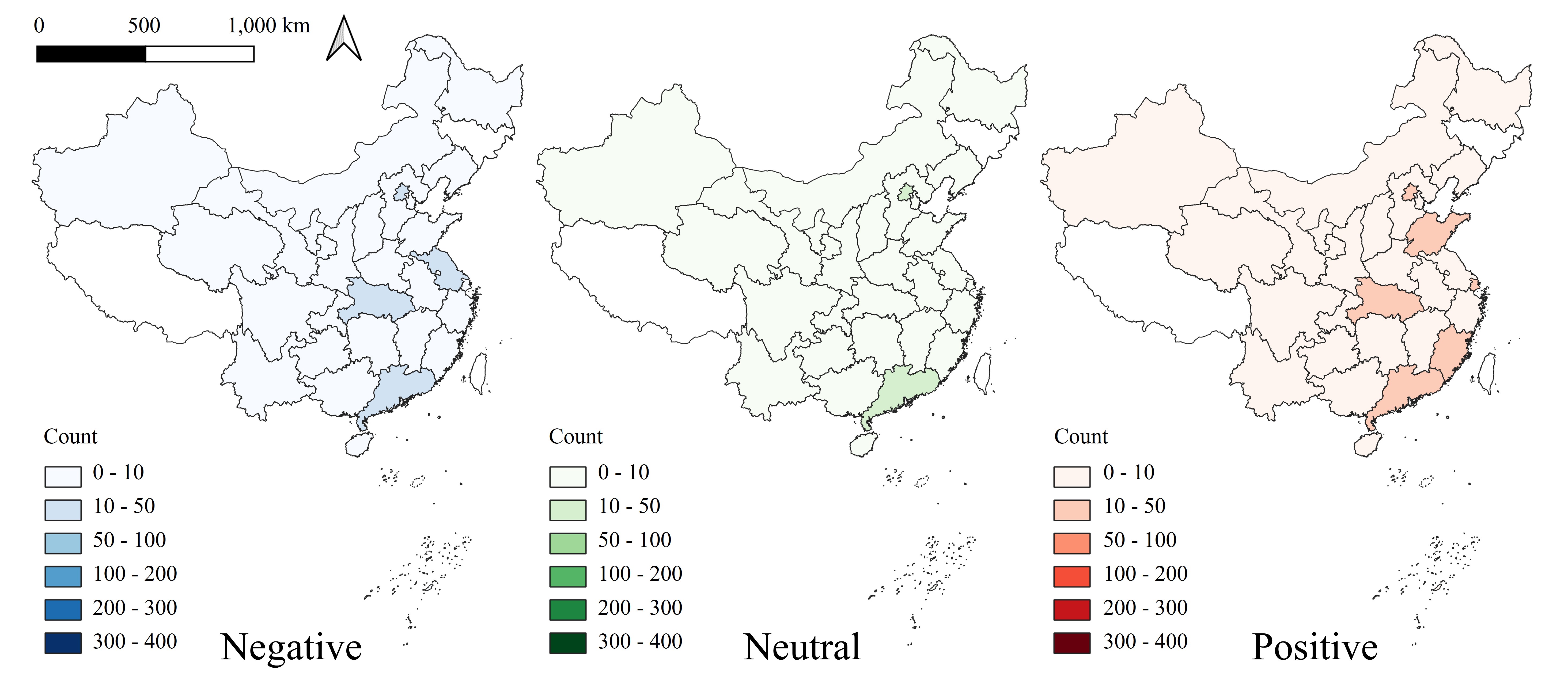}
        \caption{July 9, 2024}
        \label{fig:count_07-09}
    \end{subfigure}
    \vspace{1em} 
    \begin{subfigure}[b]{1\linewidth}
        \centering
        \includegraphics[width=0.9\linewidth]{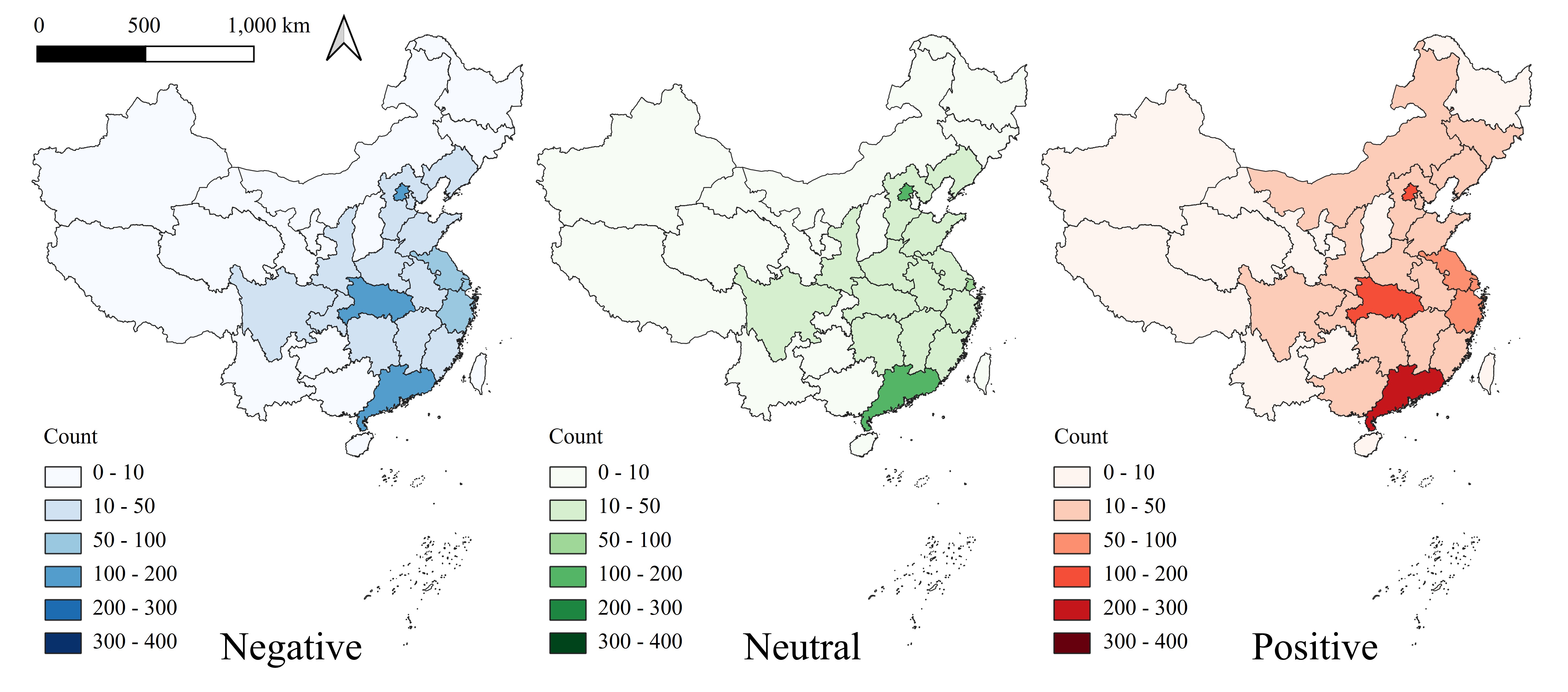}
        \caption{July 10, 2024}
        \label{fig:count_07-10}
    \end{subfigure}
\end{figure}

\begin{figure}\ContinuedFloat
    \begin{subfigure}[b]{1\linewidth}
        \centering
        \includegraphics[width=0.9\linewidth]{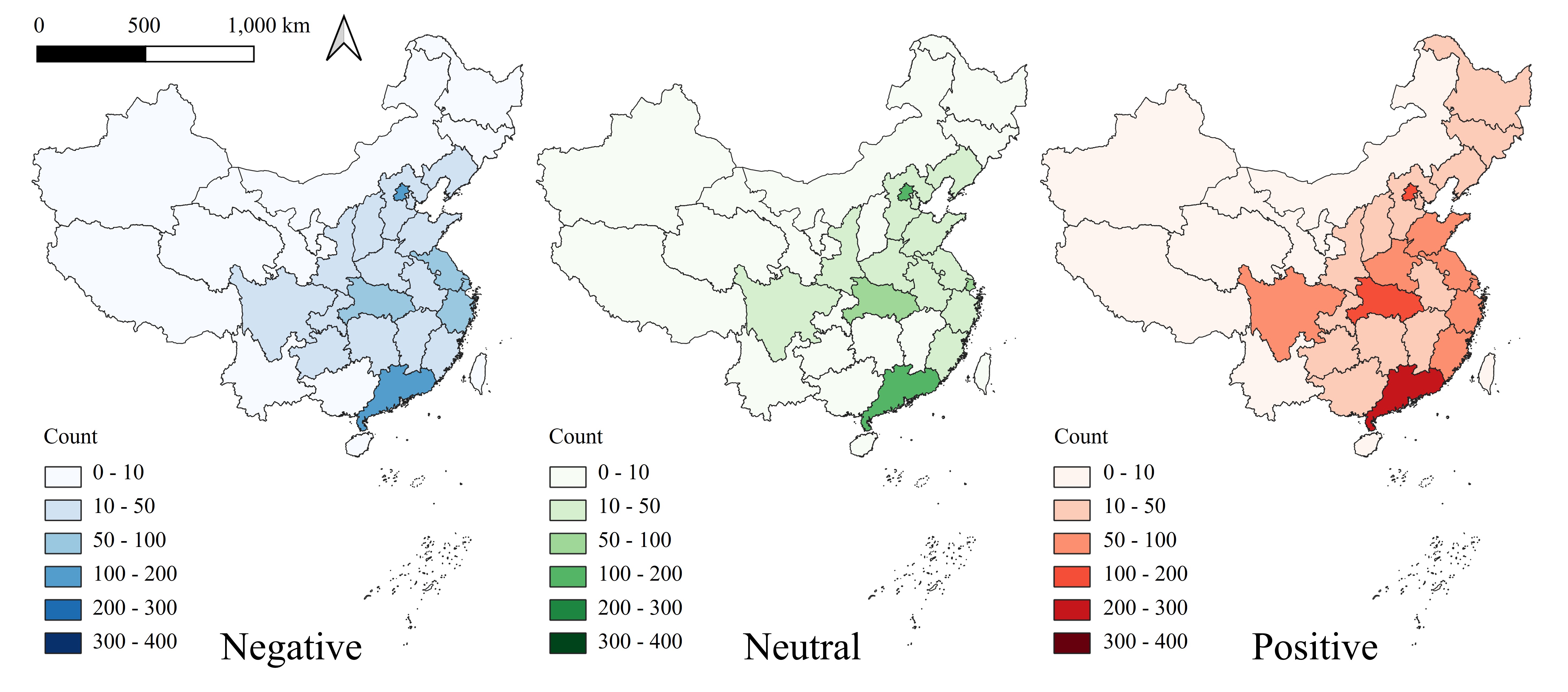}
        \caption{July 11, 2024}
        \label{fig:count_07-11}
    \end{subfigure}
    \vspace{1em} 
    \begin{subfigure}[b]{1\linewidth}
        \centering
        \includegraphics[width=0.9\linewidth]{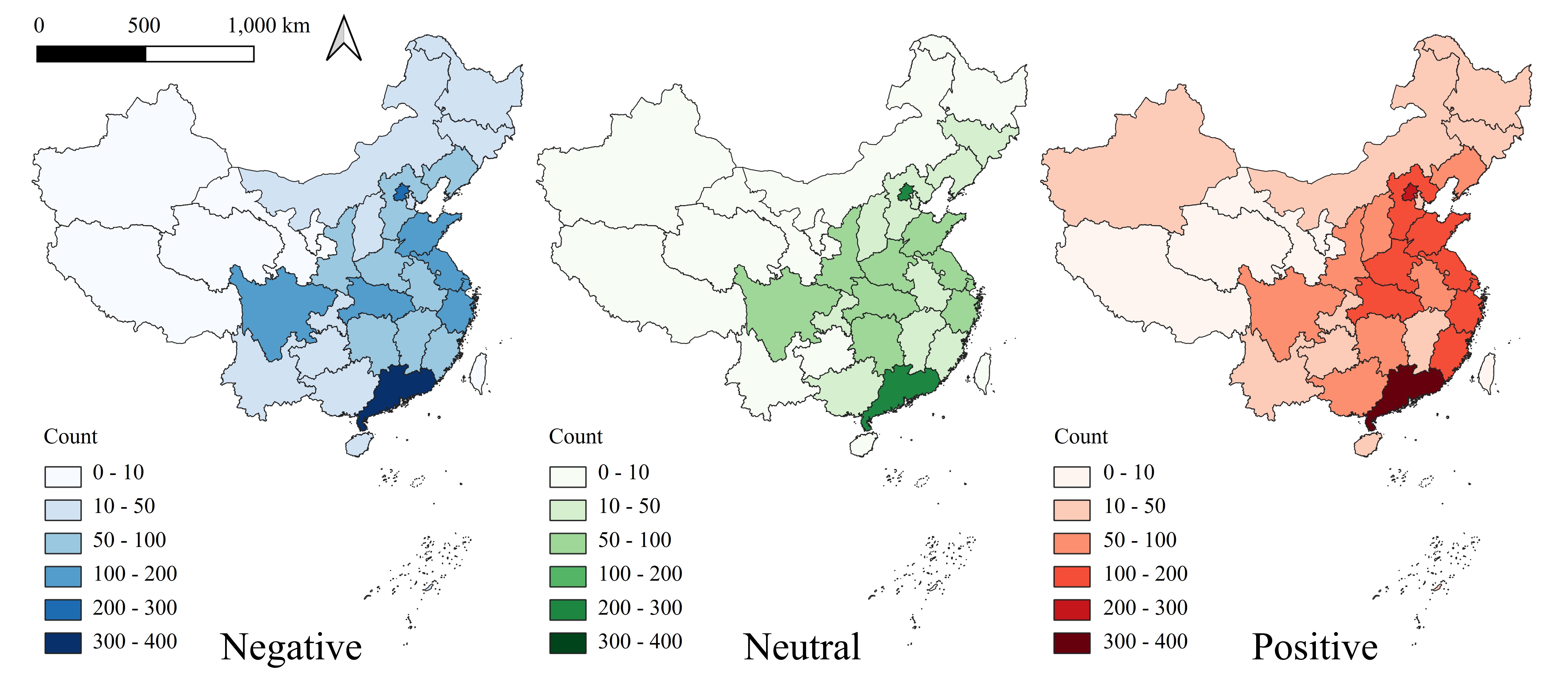}
        \caption{July 12, 2024}
        \label{fig:count_07-12}
    \end{subfigure}
    \vspace{1em} 
    \begin{subfigure}[b]{1\linewidth}
        \centering
        \includegraphics[width=0.9\linewidth]{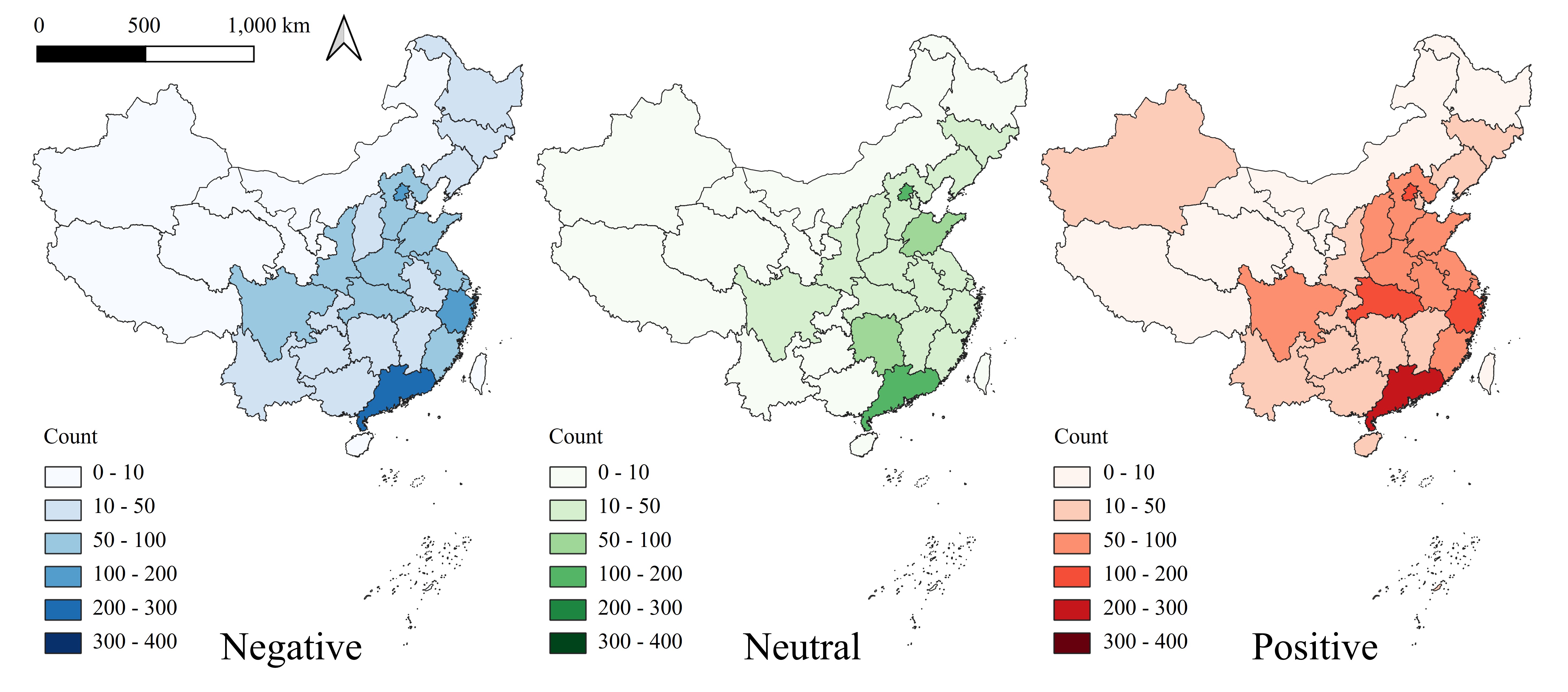}
        \caption{July 13, 2024}
        \label{fig:count_07-13}
    \end{subfigure}
\end{figure}
\begin{figure}\ContinuedFloat
    \begin{subfigure}[b]{1\linewidth}
        \centering
        \includegraphics[width=0.9\linewidth]{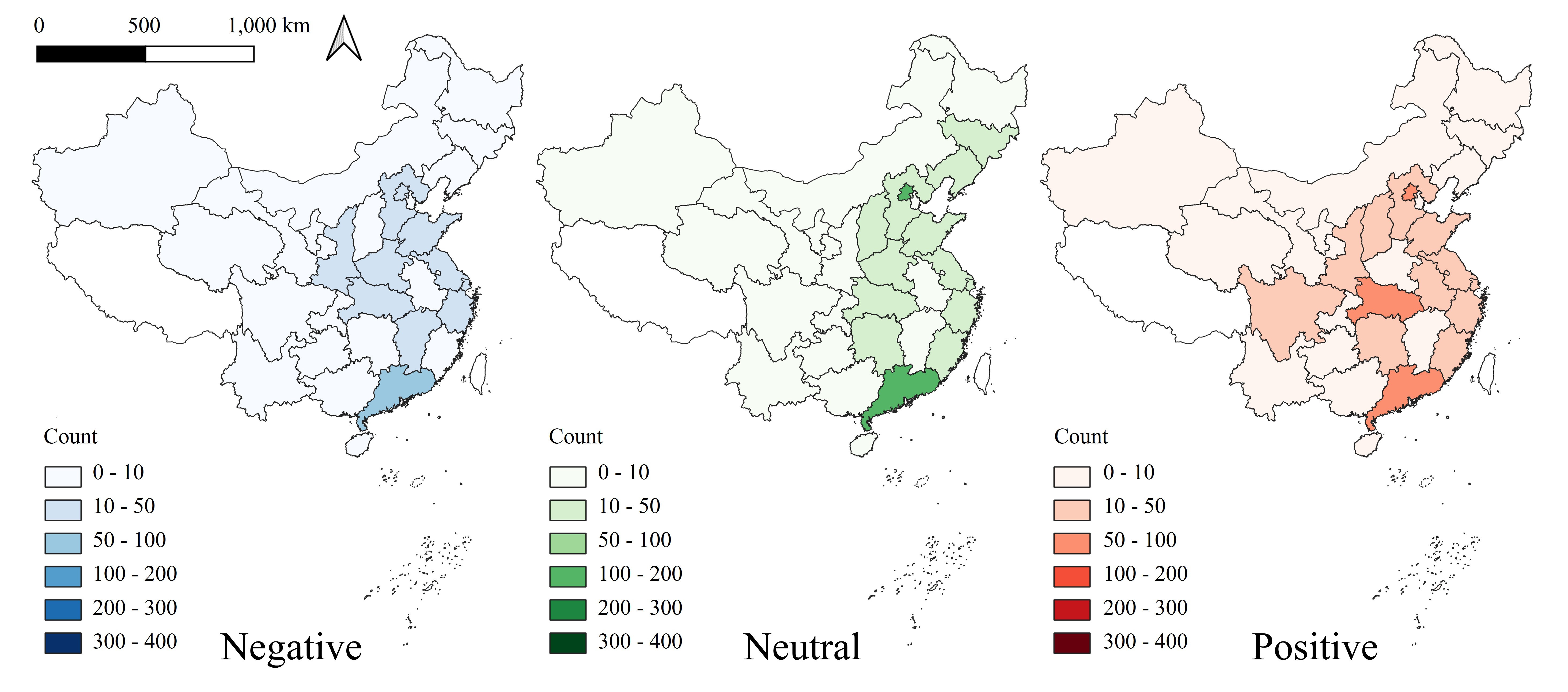}
        \caption{July 14, 2024}
        \label{fig:count_07-14}
    \end{subfigure}
    \caption{Spatial distribution of the number of negative, neutral, and positive attitudes from 8 July to 14 July 2024.}
    \label{fig:8 July to 14 July number}
\end{figure}

\begin{figure}
    \centering
    \begin{subfigure}[b]{1\linewidth}
        \centering
        \includegraphics[width=0.9\linewidth]{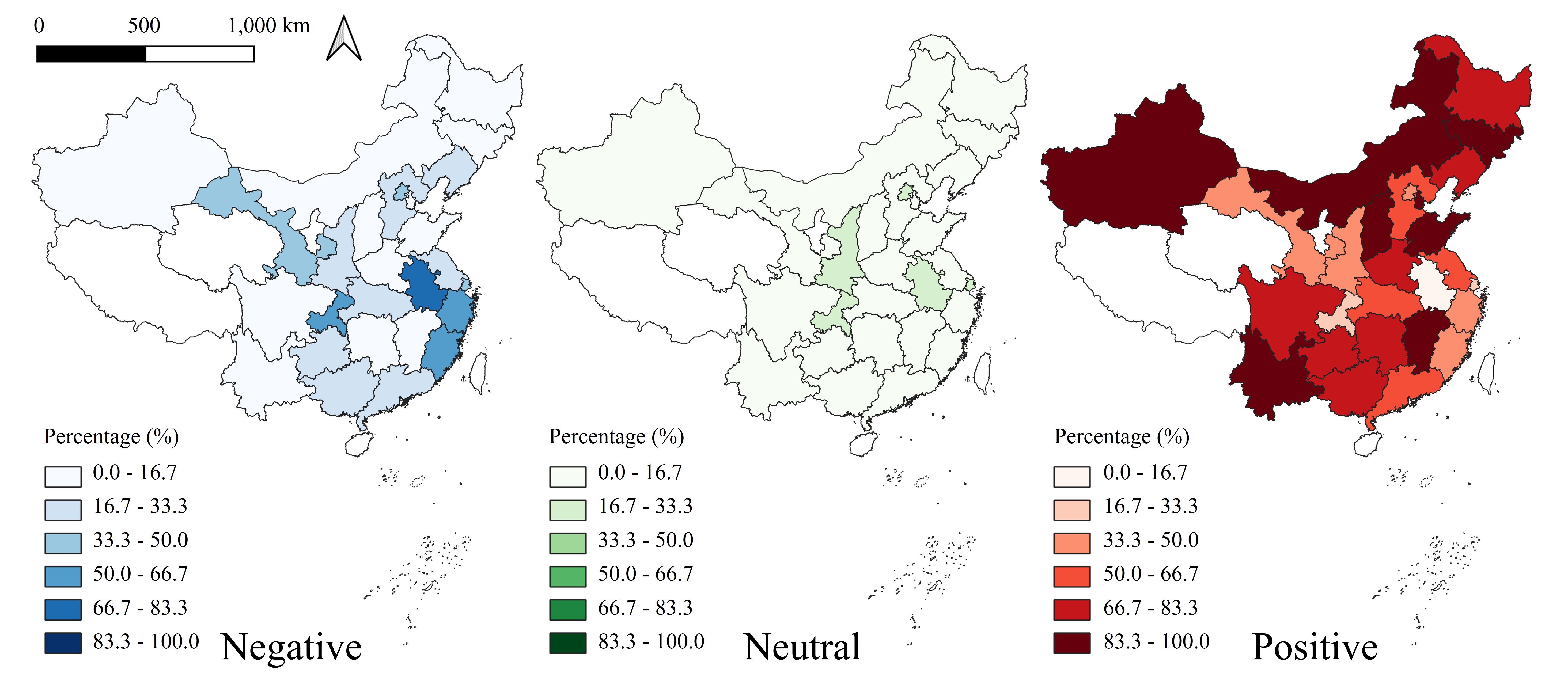}
        \caption{July 8, 2024}
        \label{fig:percentage_07-08}
    \end{subfigure}
    \vspace{1em} 
    \begin{subfigure}[b]{1\linewidth}
        \centering
        \includegraphics[width=0.9\linewidth]{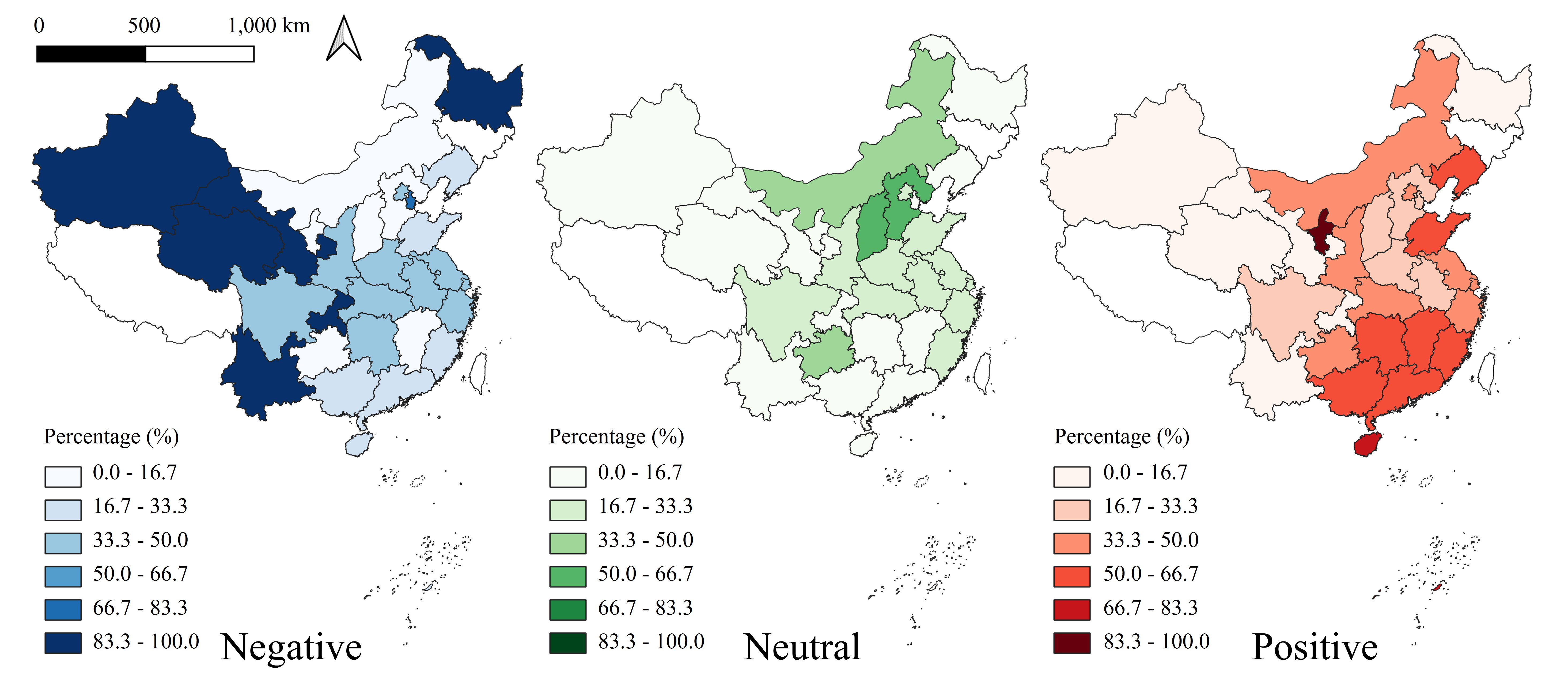}
        \caption{July 9, 2024}
        \label{fig:percentage_07-09}
    \end{subfigure}
    \vspace{1em} 
    \begin{subfigure}[b]{1\linewidth}
        \centering
        \includegraphics[width=0.9\linewidth]{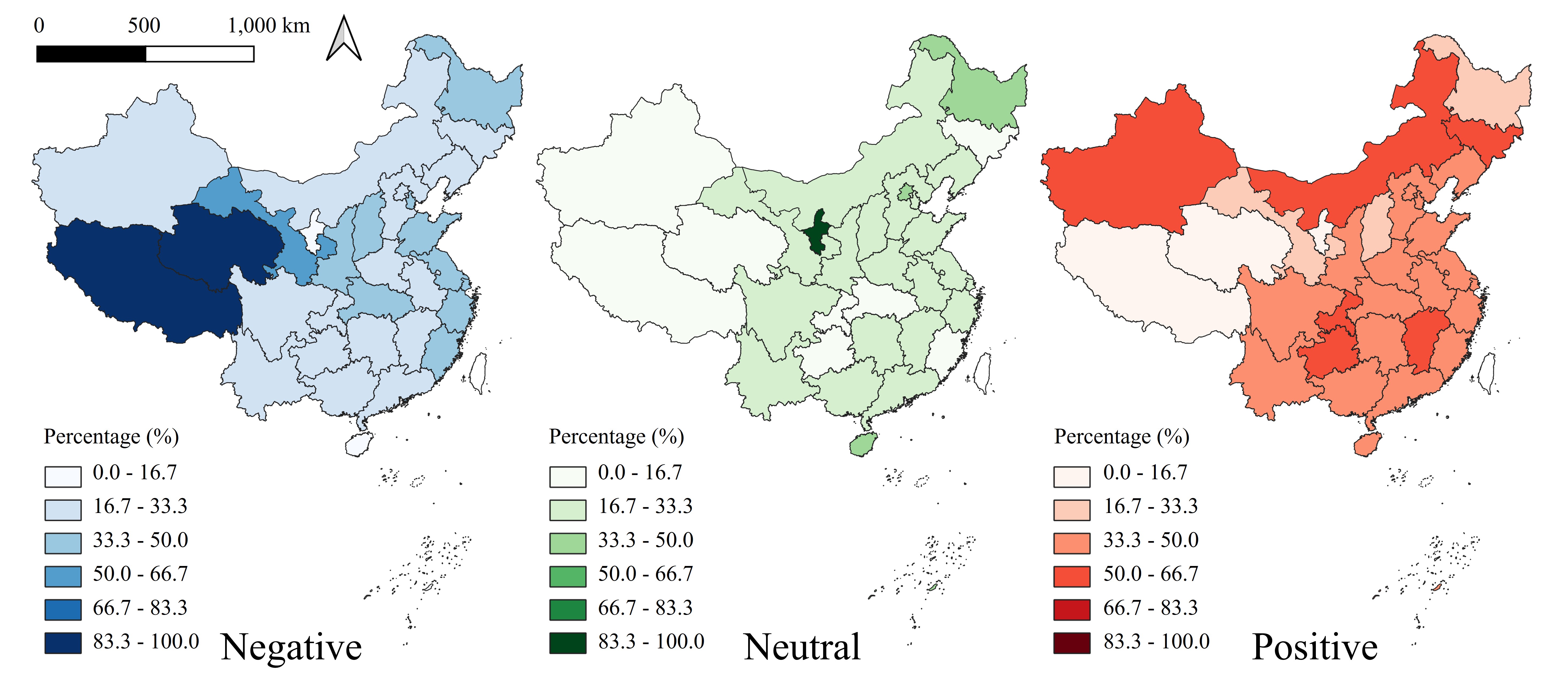}
        \caption{July 10, 2024}
        \label{fig:percentage_07-10}
    \end{subfigure}
    \label{fig:8 July to 10 July percentage}
\end{figure}

\begin{figure}\ContinuedFloat
    \centering
    \begin{subfigure}[b]{1\linewidth}
        \centering
        \includegraphics[width=0.9\linewidth]{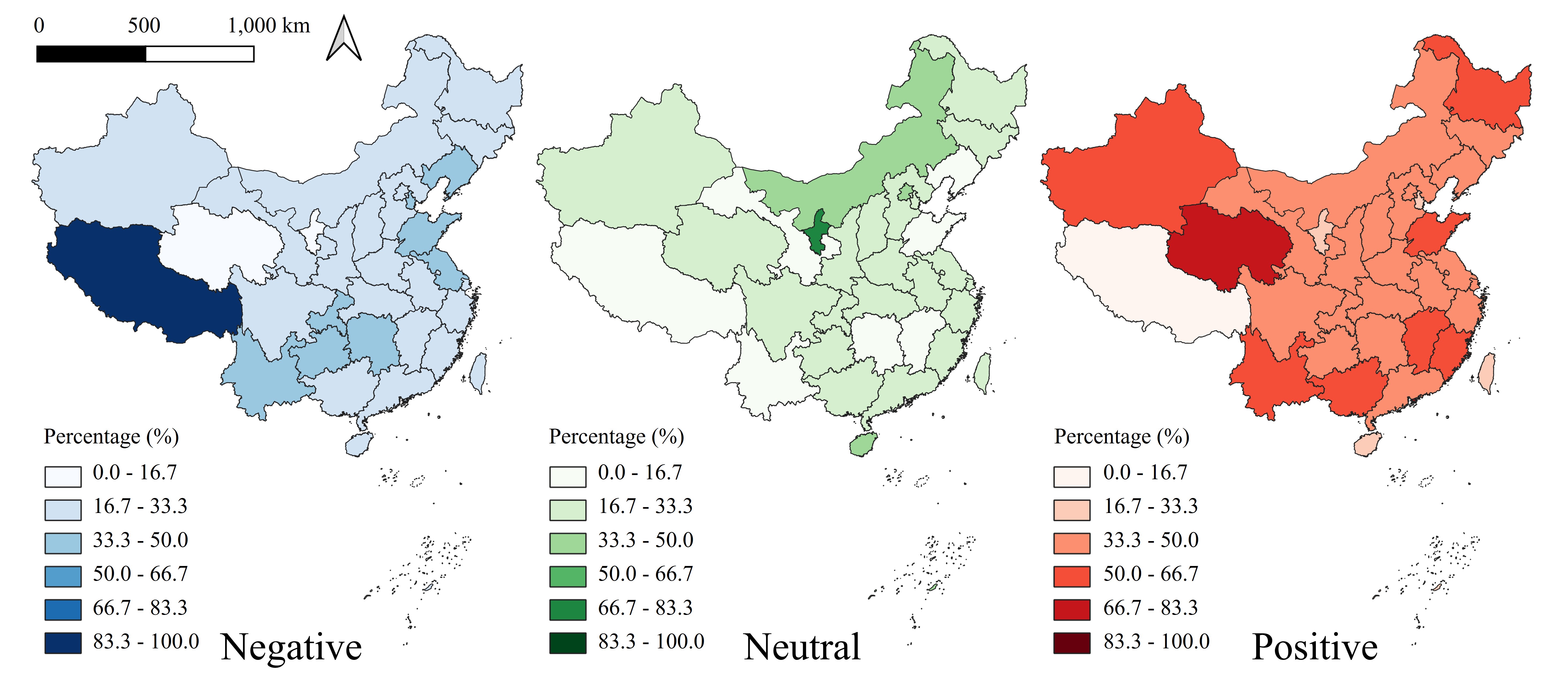}
        \caption{July 11, 2024}
        \label{fig:percentage_07-11}
    \end{subfigure}
    \vspace{1em} 
    \begin{subfigure}[b]{1\linewidth}
        \centering
        \includegraphics[width=0.9\linewidth]{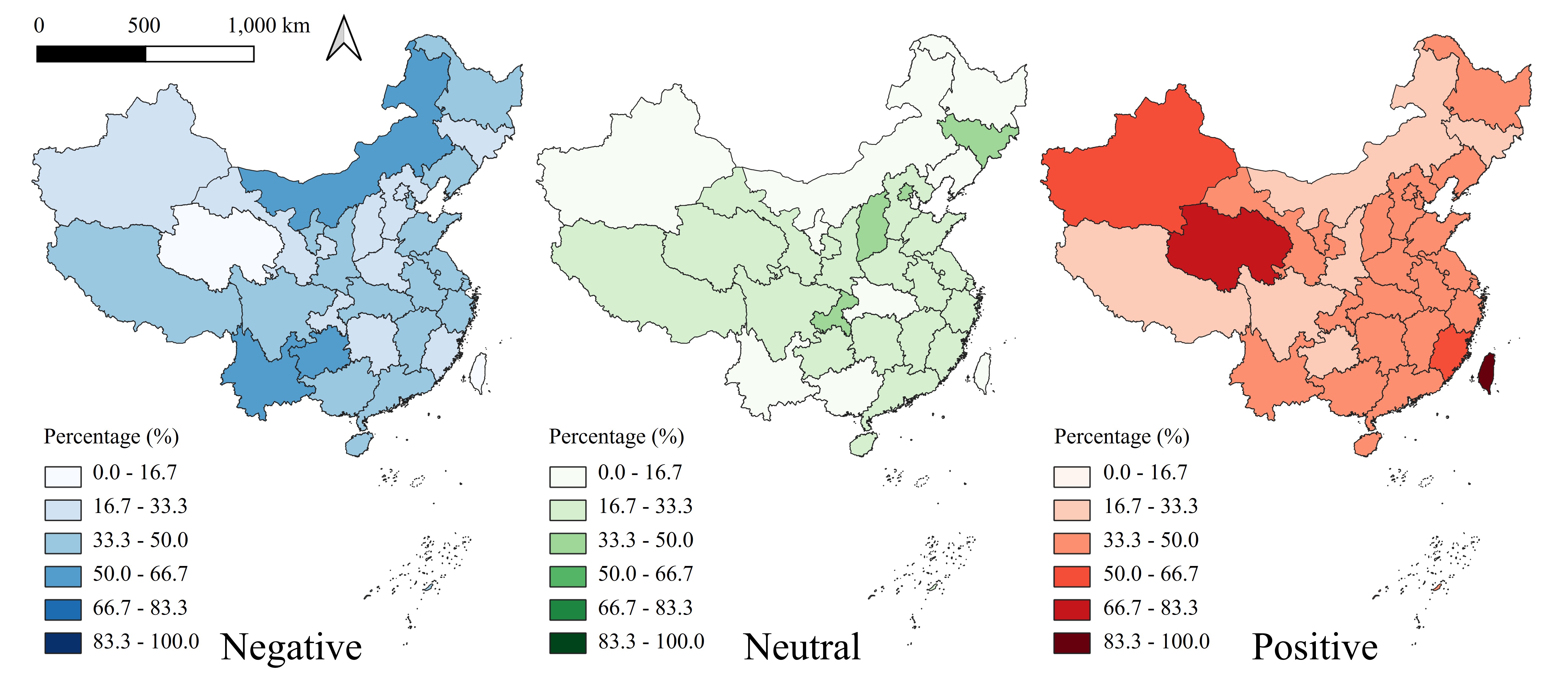}
        \caption{July 12, 2024}
        \label{fig:percentage_07-12}
    \end{subfigure}
    \vspace{1em} 
    \begin{subfigure}[b]{1\linewidth}
        \centering
        \includegraphics[width=0.9\linewidth]{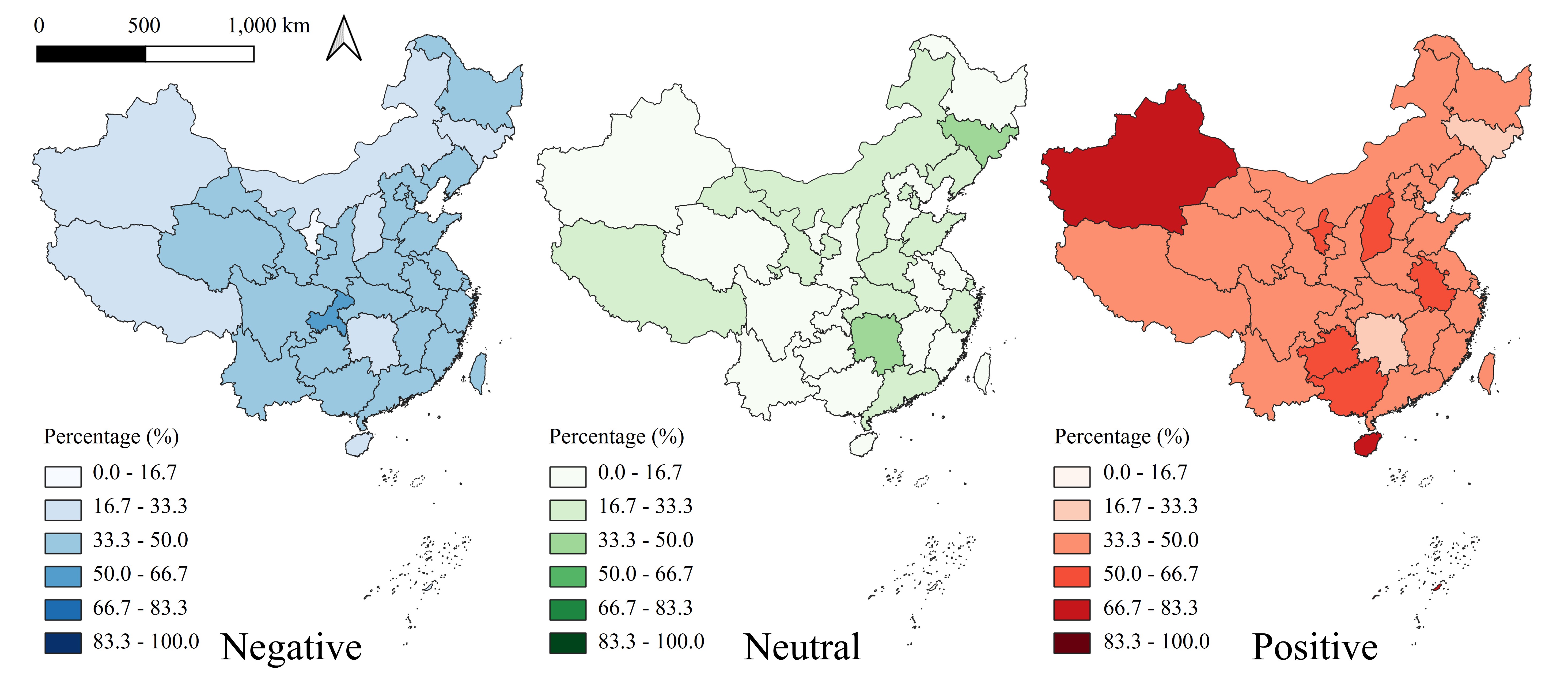}
        \caption{July 13, 2024}
        \label{fig:percentage_07-13}
    \end{subfigure}
\end{figure}
    
\begin{figure}\ContinuedFloat
    \centering
    \begin{subfigure}[b]{1\linewidth}
        \centering
        \includegraphics[width=0.9\linewidth]{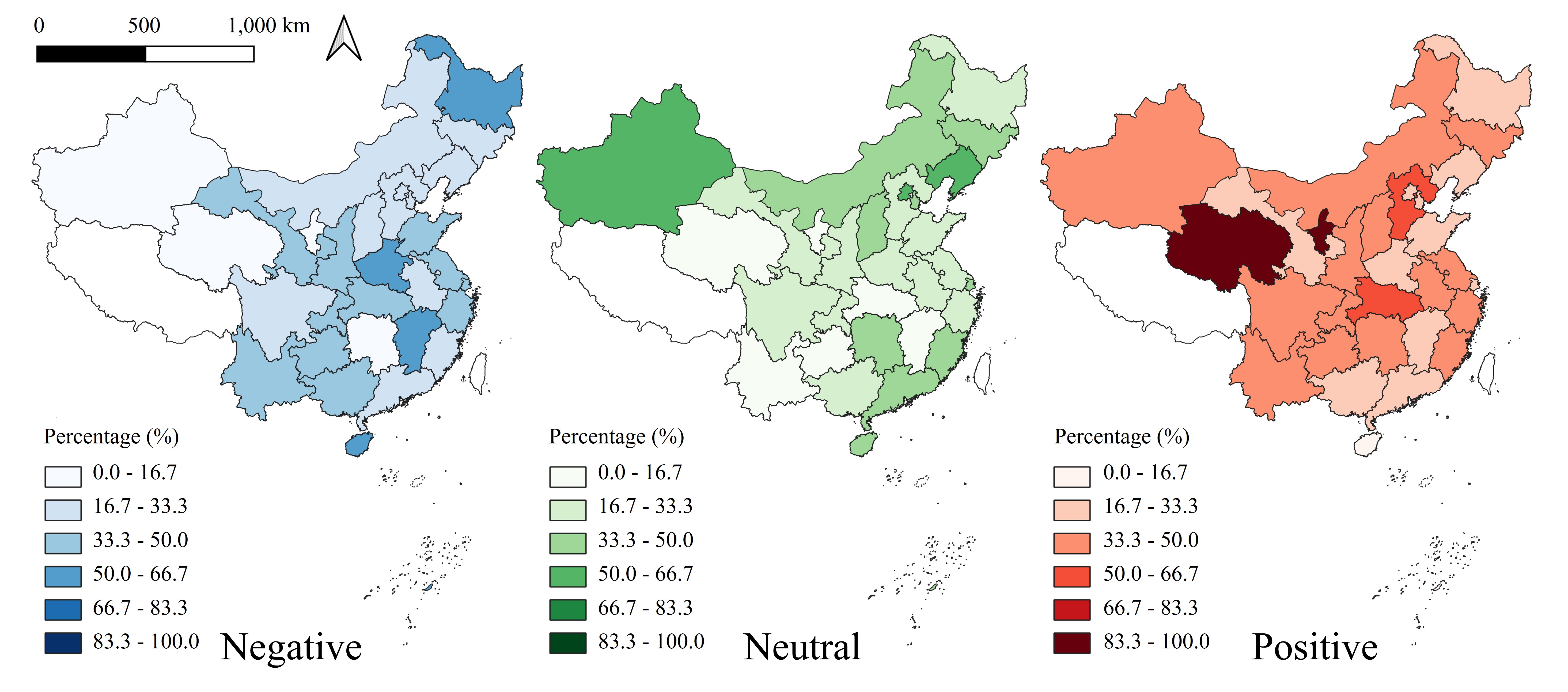}
        \caption{July 14, 2024}
        \label{fig:percentage_07-14}
    \end{subfigure}
    \caption{Spatial distribution of the percentage of negative, neutral, and positive attitudes from 8 July to 14 July 2024.}
    \label{fig:8 July to 14 July percentage}
\end{figure}

\subsection{Sentiment Analysis}

\begin{figure}
    \centering
    \begin{subfigure}{0.48\textwidth} 
        \centering
        \includegraphics[width=1\linewidth]{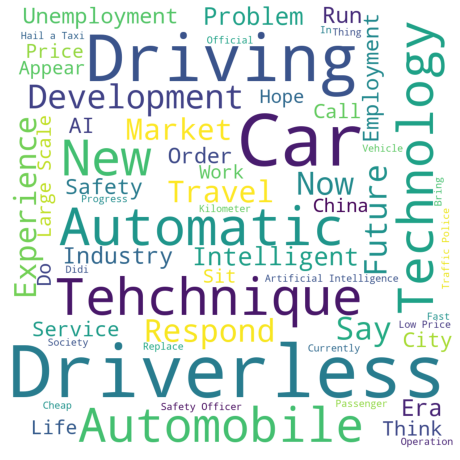}
        \caption{Most Frequent Words in the Positive Comments for Apollo Go.}
        \label{fig: Wordclouds of positive comments}
    \end{subfigure}
    \hfill 
    \begin{subfigure}{0.48\textwidth} 
        \centering
        \includegraphics[width=1\linewidth]{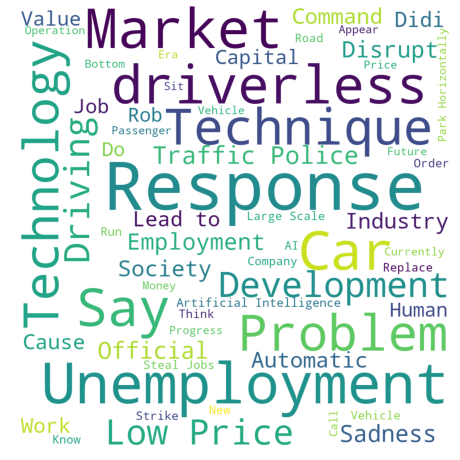}
        \caption{Most Frequent Words in the Negative Comments for Apollo Go.}
        \label{fig: Wordclouds of negative comments}
    \end{subfigure}
    \caption{Wordclouds of different attitudes.}
    \label{fig:combined_wordclouds}
\end{figure}

Word clouds (as shown in Figure~\ref{fig:combined_wordclouds}) were generated to compare the keywords in positive and negative comments for Apollo Go. Figure~\ref{fig: Wordclouds of positive comments}) shows that the words most commonly used in the positive comments for Apollo Go are “driverless”, “automatic”, “technique“, “technology“, and “development”, etc; while Figure~\ref{fig: Wordclouds of negative comments}) shows that “response”, “unemployment”, “market”, and “problem” are the high-frequency keywords in the negative comments. 


A check on the corresponding post texts reveals that most of the positive comments focused on the discussion of technology applications (such as self-driving and artificial intelligence), personal experience of Apollo Go, as well as the positive news and reports about the self-driving car company. The negative discussions focused mainly on the concerns about job displacement from driverless cars in transportation roles and the safety issues, as well as the negative news such as robotaxi breakdown.

\section{Conclusion and Discussion}\label{section5}

Autonomous ride-hailing services are rapidly growing globally, and the general public's acceptance will determine their success~\cite{wang2022investigating, jiang2021public}. This study evaluates public sentiment toward Baidu Apollo Go, a leading autonomous ride-hailing service provider in China, by analyzing social media data, particularly from Weibo. Our analysis reveals a divergence in public perceptions of shared self-driving transportation. 

Public interest in autonomous ride-hailing services varies both temporally and geographically, predominantly concentrated in China's economically developed regions. Initially, the discussions on Weibo gradually shifted from Hubei province to the provinces and cities where autonomous ride-hailing services were launched. In the later stage, those placed with no Apollo Go available, we have seen more and more people are expressing their interests, suggesting that trial runs and public testing have helped to increase public awareness of such services. Similar research has also shown that social incidents are more sensitive to the general public \cite{ding2021sentiments, jing2023listen}, indicating that effectively utilizing such cases can help expose the service to a broader audience.

As for the public comments, our word frequency analysis indicates frequent mentions and comparisons with Baidu's competitors, including Tesla and Waymo. This competitive landscape is a focal point in the discussions on Chinese social media. Positive comments predominantly highlight technological innovations and the unique experiences offered by autonomous driving, which is similar with studies conducted in the U.S.~\cite{ding2021sentiments}. In contrast, negative comments primarily express concerns about safety issues and social implications, especially job losses, while neutral comments often focus on brand-related keywords. The difference is that in other parts of the world, there is more discussion about personal privacy and data security~\cite{chen2021using}. This discrepancy may be attributed to the longer timeframe for the launch of self-driving services in those regions, allowing for a broader range of issues to surface. Interestingly, in China, some comments reflect dissatisfaction with traditional ride-hailing drivers' behaviors, such as smoking and chatting, suggesting a preference for the perceived comfort and reliability of autonomous services.

In comparison, countries like the U.S. and the U.K. have seen similar developments, with services like Waymo One and Oxbotica expanding operations in major cities. However, discussions in the U.S. tend to focus more on cities already implementing or planning to deploy these services, such as San Francisco, Phoenix, and Detroit ~\cite{jiang2021public}. This also reminds us of an advantage of using Weibo data: it contains the user's real IP address, offering a more precise measure of geographical data compared to X (Twitter), where users' locations are only inferred from their text.

Despite these insights, this study has several limitations. Firstly, the lack of social demographic information, such as age and income, limits our understanding of the factors influencing public attitudes. Existing studies have shown that these attributes are critical in shaping public perceptions~\cite{wang2022investigating}. Secondly, the data volume and time span are relatively small, with discussions on social media platforms heating up late in the study period. In contrast, other similar studies typically observe data continuously for about three months~\cite{chen2021using, jiang2021public}, offering a more comprehensive view of public sentiment over time.

Moving forward, it would be beneficial for future research to explore the impact of various demographic characteristics on public attitudes and to extend the observation period for a more comprehensive analysis. It could also helpful in combining social media data with other traditional datasets. Such findings can serve as crucial references for planners, policymakers, and service providers aiming to promote autonomous ride-hailing services. Effective strategies might include enhancing public outreach, conducting trial rides, and refining legal frameworks to mitigate potential negative impacts, such as job losses or liability issues in accidents. With these measures, the future of autonomous ride-hailing services looks promising.



\bibliographystyle{unsrt}  
\bibliography{refs}  
\end{document}